\documentclass[final,5p,times,twocolumn,nopreprintline]{elsarticle}

\usepackage{amssymb}
\usepackage{amsmath}

\usepackage{multirow} 
\usepackage{hyperref}
 \usepackage{booktabs}

\journal{Medical Image Analysis}

\begin{document}

\begin{frontmatter}


\tnotetext[github]{\href{https://github.com/ShadowTwin41/generative_networks}{The code is available on GitHub: \\https://github.com/ShadowTwin41/generative\_networks}
}

\title{Enhancing Privacy: The Utility of Stand-Alone Synthetic CT and MRI for Tumor and Bone Segmentation}

\author[Uminho,IKIM,RWTHinformatics,RWTHsurgery]{André Ferreira \corref{cor1}} 
\author[RWTHinformatics,RWTHsurgery]{Kunpeng Xie}
\author[Freiburg]{Caroline Wilpert}
\author[Uminho]{Gustavo Correia}
\author[RWTHradio]{Felix Barajas Ordonez}
\author[UminhoICVS]{Tiago Gil Oliveira}
\author[RWTHradio]{Maike Bode}
\author[RWTHradio]{Robert Siepmann}
\author[RWTHsurgery]{Frank Hölzle} 
\author[RWTHinformatics]{Rainer Röhrig}
\author[IKIM]{Jens Kleesiek}
\author[RWTHradio]{Daniel Truhn}
\author[IKIM,CCCE]{Jan Egger}
\author[Uminho]{Victor Alves}
\author[RWTHinformatics,RWTHsurgery]{Behrus Puladi}

\affiliation[Uminho]{organization={Center Algoritmi / LASI},
            addressline={University of Minho}, 
            city={Braga},
            postcode={4710-057}, 
            country={Portugal}}
\affiliation[IKIM]{organization={Institute for AI in Medicine (IKIM)},
            addressline={University Hospital Essen (AöR)}, 
            city={Essen},
            postcode={45131}, 
            country={Germany}}
\affiliation[RWTHinformatics]{organization={Institute of Medical Informatics},
            addressline={University Hospital RWTH Aachen}, 
            city={Aachen},
            postcode={52074}, 
            country={Germany}}
\affiliation[RWTHsurgery]{organization={Department of Oral and Maxillofacial Surgery},
            addressline={University Hospital RWTH Aachen}, 
            city={Aachen},
            postcode={52074}, 
            country={Germany}}
\affiliation[Freiburg]{organization={Department of Diagnostic and Interventional Radiology, University Medical Center Freiburg, Faculty of Medicine},
            addressline={University of Freiburg}, 
            city={Freiburg},
            country={Germany}}
\affiliation[RWTHradio]{organization={Department of Diagnostic and Interventional Radiology},
            addressline={University Hospital RWTH Aachen}, 
            city={Aachen},
            postcode={52074}, 
            country={Germany}}
\affiliation[UminhoICVS]{organization={Life and Health Sciences Research Institute (ICVS), School of Medicine},
            addressline={University of Minho}, 
            city={Braga},
            postcode={4710-057}, 
            country={Portugal}}     
\affiliation[CCCE]{organization={Cancer Research Center Cologne Essen (CCCE)},
            addressline={University Medicine Essen, Hufelandstraße 55}, 
            city={Essen},
            postcode={45147}, 
            country={Germany}}
\cortext[cor1]{Corresponding author: 
  email: id10656@alunos.uminho.pt;}
\begin{abstract}
AI requires extensive datasets, while medical data is subject to high data protection. Anonymization is essential, but poses a challenge for some regions, such as the head, as identifying structures overlap with regions of clinical interest. Synthetic data offers a potential solution, but studies often lack rigorous evaluation of realism and utility.
Therefore, we investigate to what extent synthetic data can replace real data in segmentation tasks.
We employed head and neck cancer CT scans and brain glioma MRI scans from two large datasets. Synthetic data were generated using generative adversarial networks and diffusion models. We evaluated the quality of the synthetic data using MAE, MS-SSIM, Radiomics and a Visual Turing Test (VTT) performed by 5 radiologists and their usefulness in segmentation tasks using DSC.
Radiomics indicates high fidelity of synthetic MRIs, but fall short in producing highly realistic CT tissue, with correlation coefficient of 0.8784 and 0.5461 for MRI and CT tumors, respectively.
DSC results indicate limited utility of synthetic data: tumor segmentation achieved DSC=0.064 on CT and 0.834 on MRI, while bone segmentation a mean DSC=0.841. Relation between DSC and correlation is observed, but is limited by the complexity of the task. VTT results show synthetic CTs’ utility, but with limited educational applications. Synthetic data can be used independently for the segmentation task, although limited by the complexity of the structures to segment. Advancing generative models to better tolerate heterogeneous inputs and learn subtle details is essential for enhancing their realism and expanding their application potential.
\end{abstract}


\end{frontmatter}


\section{Introduction}
\label{Introduction}

The most important breakthroughs in Artificial Intelligence (AI) are taking place in areas where large and easily accessible amounts of data are available. It is widely recognized that AI solutions require large datasets for training, especially those based on Deep Learning (DL) \citep{wang2020big}. Medical images are acquired in different centers every day, increasing the amount of data available for various tasks. Due to data protection regulations, this data is often underutilized and eventually forgotten within medical facilities \citep{ahmed2023systematic}. The sensitivity of data results in stricter regulations that can limit access to this information. Regions such as the head and face cannot be anonymized through conventional methods such as defacing, as the area of clinical interest coincides with the region requiring removal for privacy. These constraints make it difficult to create comprehensive medical datasets, limiting researchers from fully exploiting the potential of AI in healthcare \citep{muschelli2019recommendations}. Therefore, creating a large dataset of sensitive information, such as head and neck, with assured quality and sufficient variability that faithfully represents the real population is a very time-consuming and expensive task as it legally requires obtaining informed consent from each individual \citep{kadam2017informed,howe2020ethical}. 

\begin{figure*}[t]
\centering
\includegraphics[width=0.9\textwidth]{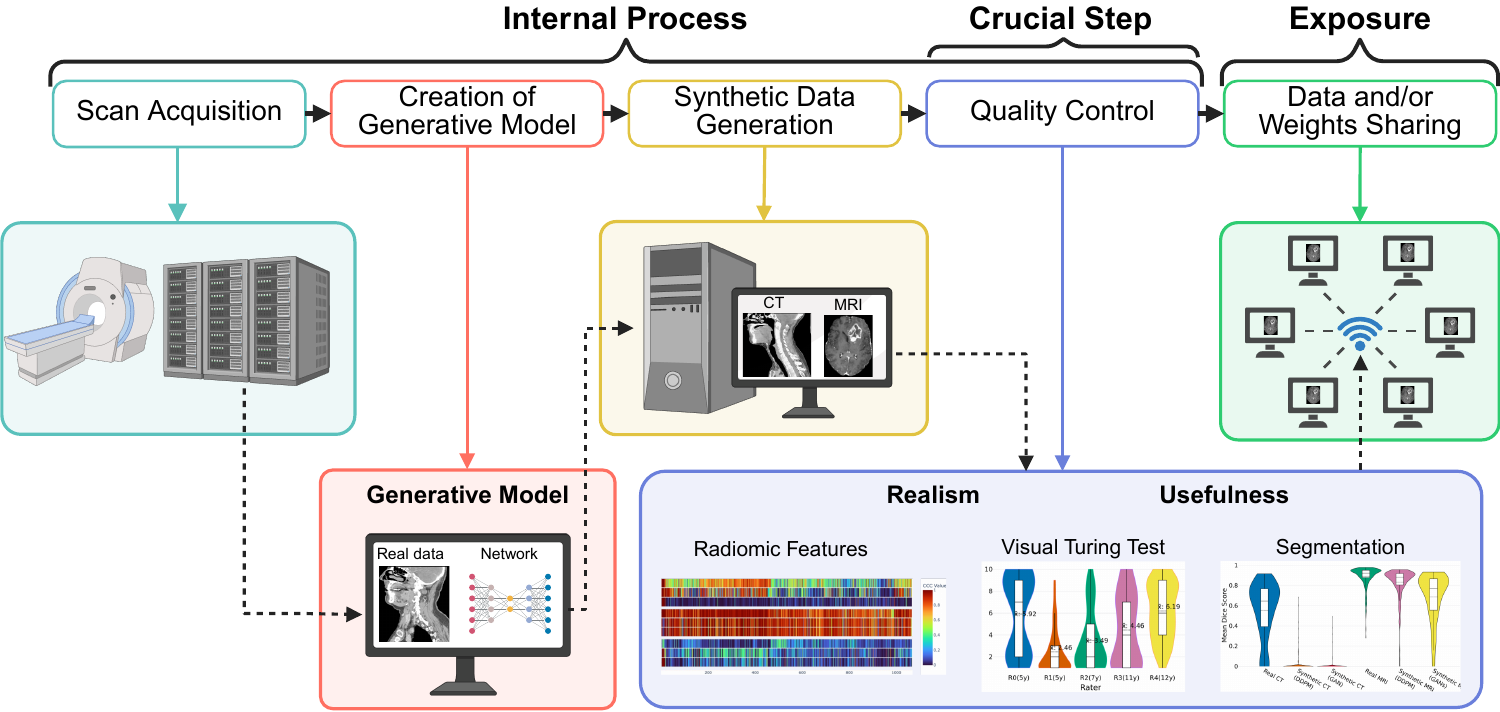}
\caption{Study Workflow: The process began with the collection and organization of MRI and CT scans, which were used to train generative models, specifically Generative Adversarial Networks (GANs) and Denoising Diffusion Probabilistic Models (DDPMs). Subsequently, synthetic datasets were generated and subjected to systematic evaluation to assess their realism and potential utility. Finally, the datasets and/or the weights trained for the downstream task (segmentation) can be prepared for secure sharing, if their realism and usefulness are proven. Created in BioRender. Xie, K. (2025) https://BioRender.com/n5bgr3d.}
\label{fig:overall}
\end{figure*}

\begin{figure}[t]
\centering
\includegraphics[width=0.49\textwidth]{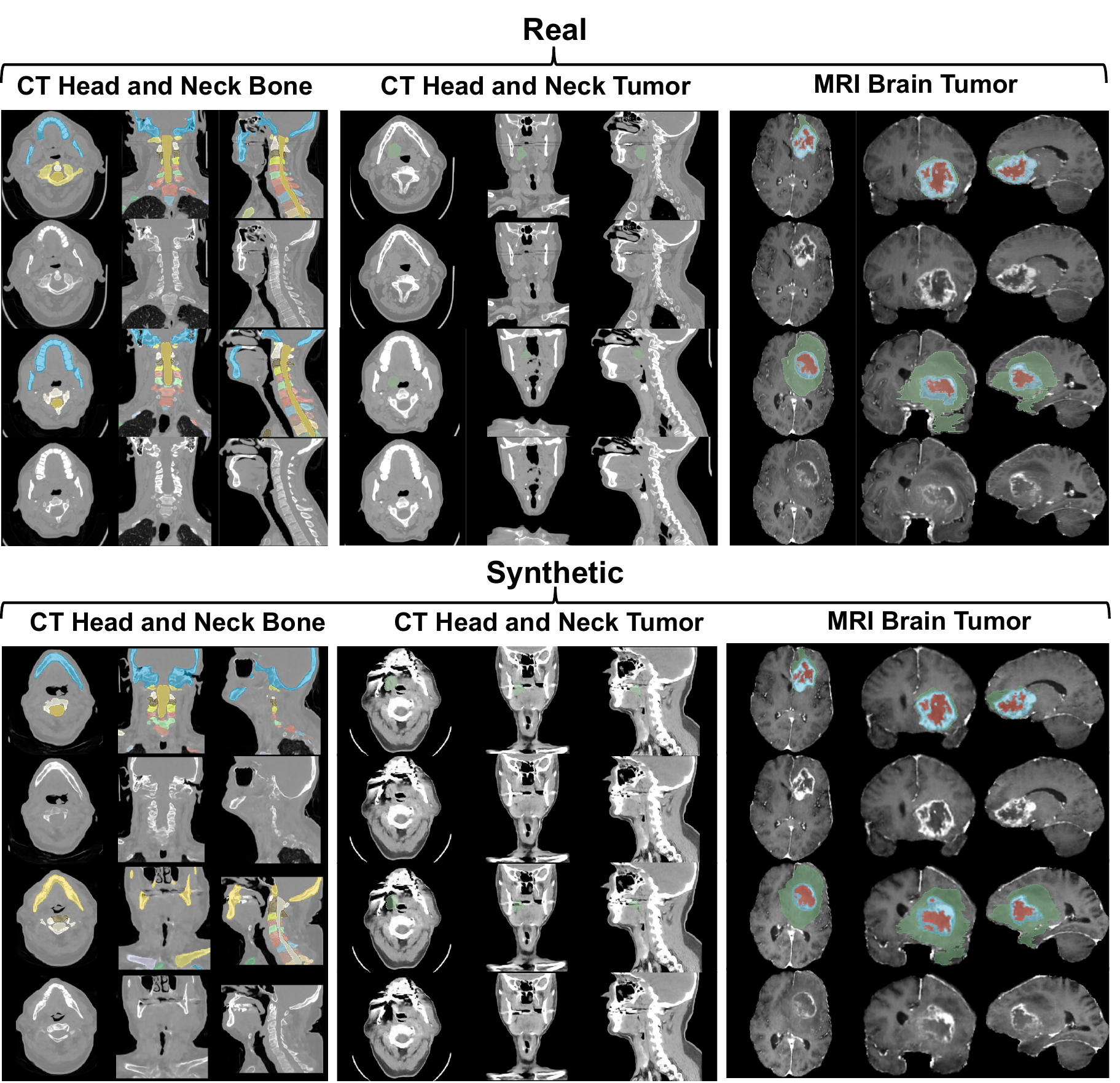}
\caption{Visual comparison between real and synthetic samples for both tumor and bone segmentation tasks. Top row: Real cases. Bottom row: Synthetic cases.}
\label{fig:overall_synth}
\end{figure}

Synthetic data and Federated Learning (FL) have emerged as two key strategies for leveraging protected data in inter-institutional research while ensuring that sensitive information remains confidential and secure \citep{blanco2021achieving,mendes2025synthetic}. FL \citep{yang2019federated} is a strategic approach that facilitates the decentralized training of Machine Learning (ML) models using remotely hosted datasets. However, it is crucial to ensure that the weights are securely shared with the main server and that they are resistant to attacks in case of a breach in communication or disclosure of the trained models. Additionally, the necessary infrastructure and the requirement for agreement among multiple institutions increase the complexity of this approach \citep{kandati2023security,zerka2020systematic}. The generation of synthetic data does not have these constrains. It can be created locally and shared or used for a downstream task. When produced by generative models, it is in principle anonymous and follow the real data distribution, which makes such approach a good solution for data sharing, as well as enable privacy-preserving synthetic data generation for education \citep{yoon2020anonymization,kazerouni2023diffusion}.

The application of synthetic data for downstream tasks has been explored in the literature \citep{murtaza2023synthetic,ferreira2024gan}. Often, it is used as a supplementary tool to enhance downstream tasks, such as through pre-training or as a data augmentation technique. \cite{liu2023utilizing} pre-trained DL models on synthetic data and fine-tuned them for various downstream tasks, i.e., classification, segmentation, and object detection. The amount of real data necessary to achieve good performance was studied, concluding that pre-training on synthetic data allows for the use of a smaller amount of real data. \cite{bhattacharya2023generating} generated synthetic Computed Tomography (CT) scans from X-rays to improve the classification of normal and abnormal images. 
The use of synthetic CT scans alongside real X-rays demonstrated improvements in classification performance. Data augmentation can also serve to improve fairness, balance datasets and increase the total amount of data available for training \citep{ferreira2024gan}. 
\cite{blow2025data} trained a diffusion models to generate synthetic tabular data for reweighting samples, which improved binary classification. \cite{ferreira2024we} utilized conditional Generative Adversarial Networks (cGANs) to generate synthetic Magnetic Resonance Imaging (MRI) of brain tumors in healthy regions, thereby increasing the variability of tumors for training a segmentation model. The inclusion of more diverse tumors resulted in a more robust segmentation model. \cite{hosseini2025synthetic} utilized diffusion models to generate synthetic chest X-rays, ophthalmology optical coherence tomography, and histopathology cases. Classification models were then trained using either only real data or only synthetic data, demonstrating that similar results can be achieved with both types of data. 

A thorough evaluation of data features is crucial. While synthetic data may seem realistic, it often lacks important characteristics. In medical imaging, Radiomic features are commonly examined. \cite{yuan2023comprehensive} conducted a detailed analysis comparing synthetic and real CT images for nasopharyngeal carcinoma. They employed metrics such as Mean Absolute Error (MAE), Root Mean Squared Error (RMSE), Structural Similarity Index (SSIM), and Peak Signal-to-Noise Ratio (PSNR) to assess the synthetic images. However, they placed significant emphasis on the Concordance Correlation Coefficient (CCC) of Radiomic features between real CT and synthetic CT. The findings indicate that most of the features demonstrate poor similarity. \cite{chen2021generative} also addressed the challenge of replicating Radiomic features, finding that generative models can enhance noisy CT scans and improve CCC. 

Visual Turing Test (VTT) is also a strong and crucial strategy for evaluating the realism of synthetic data \citep{ferreira2024gan}. However, obtaining a sufficient number of assessments for meaningful results necessitates several hours of focused attention from radiologists \citep{rozenshtein2024us}. Additionally, establishing the necessary conditions to conduct these evaluations further complicates the process. Despite these challenges, such evaluations are essential for translating these methodologies into clinical practice. \cite{jang2023image} trained two progressively growing GAN with normal and abnormal high-resolution 2D chest X-ray images. Six radiologists were selected to perform a binary (real or synthetic) VTT. The results show that the experts were not able to distinguish all synthetic cases, although more experienced radiologists were able to distinguish abnormal cases.
In \cite{khader2023denoising}, two radiologists performed a VTT on volumetric scans generated by a 3D Denoising Diffusion Probabilistic Models (DDPM) and vector-quantized GAN, rating image quality, slice consistency and anatomical correctness. Overall realism was high, though minor anatomical errors and slice inconsistencies were noted.

The literature typically focuses on the use of synthetic data for model pre-training and/or data augmentation, lacking in-depth and systematic studies regarding the use of synthetic data alone for downstream tasks. Real-world datasets often contain subtle patterns, variations, and edge cases that are challenging to model and reproduce synthetically \citep{murtaza2023synthetic,koetzier2024generating,kokosi2022synthetic}. Synthetic data could be used to improve privacy, but this lack of rigorous studies makes it unreliable to use it in crucial points where real data cannot be used. 

Using only synthetic data would improve privacy even further than conventional methods. To this end, we systematically evaluated the feasibility of using solely synthetic data and proposed several techniques to ensure its realism and utility. Accordingly, we focus on the crucial aspect of our research, highlighted in blue in Fig. \ref{fig:overall}, i.e. the study of the quality of synthetically generated CT and MRI scans using MAE, Multi-Scale Structural Similarity Index Measure (MS-SSIM), Radiomics and a VTT performed by 5 radiologists and
their usefulness in segmentation of tumors and bone structures using Dice Similarity Coefficient (DSC). The two leading approaches for synthetic data generation cGANs and conditional DDPMs (cDDPMs) are used \citep{oulmalme2025systematic}. Our objective is to evaluate conditional approaches that facilitate the automatic generation of synthetic data and corresponding ground truth, thereby minimizing the need for specialist intervention and reducing their workload. We also employ two key and inherently time-consuming imaging modalities, MRI and CT, which are routinely used for diagnosis and treatment planning, as well as the segmentation task which is an essential yet labor-intensive task in the medical field, underscoring the need for high-quality results \citep{egger2013gbm}. Fig. \ref{fig:overall_synth} illustrates a comparison between samples from the real dataset and synthetically generated with the respective segmentations for each task.

\section{Material and Methods}
\label{sec:material_methods}
\subsection{Preparation of real CT and MRI datasets}
\label{subsec:ct_mri_datasets}
The real and synthetic datasets were formally defined as $modality_{real}^{n}$, where $modality$ stands for the modality (CT or MRI), $real$ for reality (real or fake), and $n$ for the number of cases.
\subsubsection{CT Head and Neck cancer dataset}
\label{subsubsec:ct_dataset}
1.355 cases were extracted from four datasets of public databases: Head-Neck-PET-CT \citep{vallieres2017Radiomics}, TCGA-HNSC \citep{Zuley2016}, CPTAC-HNSCC \citep{CPTAC2018}, and HEAD-NECK-RADIOMICS-HN1 \citep{aerts2014decoding, WeeDekker2019}, all from  The Cancer Imaging Archive (TCIA) \citep{clark2013cancer}. 1.258 cases were selected under the supervision of B.H. and K.X., with the remaining cases discarded due to poor quality or lack of complete head and/or neck regions. 285 cases did not have the respective tumor segmentation, but these were also included due to the small size of the dataset. All cases were interpolated to a uniform resolution of $1\mathrm{mm}^3$ using B-spline interpolation. The Region Of Interest (ROI) in this study was the head and neck area. Therefore, a bounding box detector\footnote{https://github.com/Project-MONAI/tutorials/tree/main/detection}  was trained to automatically crop all cases. It was ensured that all cases have a smaller or equal shape of 256×256×256 after cropping.

For the tumor segmentation task, the intensity values of the data were clipped within the range of [-200, 200] Hounsfield Units (HU), as the HU values corresponding to tumors typically fall within this interval and allow a clear distinction between soft tissue, tumor and hard tissue. A range of [-1000, 1000] HU was selected for the bone segmentation in order to better distinguish between bone density and surrounding tissue. For both tasks, these values are scaled linearly between [-1,1].

The portion of the dataset containing the segmentation labels was randomly split into 80\% (n=778) for training ($CT_{real}^{778}$) and 20\% (n=195) for testing ($C_{real}^{195}$). The 285 without segmentations were then added to the training cases ($778+285=CT_{real}^{1063}$ ). The test cases are hidden for all training processes to ensure a fair assessment. Some examples of real cases can be seen in \ref{app:Real_CT_scans} and \ref{app:Real_CT_scans_bone}.

\subsubsection{MRI Brain  glioblastoma dataset (BraTS2023)}
\label{subsubsect:brats_dataset}

The MRI dataset comes from the Brain Tumor Segmentation (BraTS) challenge \citep{baid2021rsna}. 1.251 contrast-enhanced T1 scans (T1c) with the corresponding segmentations were used. This dataset is already labeled, skull-striped, interpolated to the same resolution ($1mm^3$) and all cases have the same shape 240×240×155. Shapes of 256×256×256 were created by padding with the value 0.

The data was normalized by clipping between the quantiles 0.001 and 0.999 and linearly to the range [-1, 1]. The labels were split into three regions, the Enhancing Tumor (ET), the Tumor Core (TC) composed by the NeCRotic tumor core (NCR), and Peritumoral EDema (ED), and the whole tumor (WT) composed by the three segmentations (NCR+ED+ET). Approximately 80\% of the dataset was randomly selected for training ($MRI_{\text{real}}^{1000}$), and the remaining 20\% reserved for testing ($MRI_{\text{real}}^{251}$). Some examples can be seen in \ref{app:Real_MRI_scans}.

\subsection{Preparation of synthetic CT and MRI datasets}
\label{subsec:preparation_synth_CT_MRI}
The generative models were formally denoted as $Network_{condition\_mode}^{range}$, where $Network$ refers to the network type and can be one of $3D_{cWDM}$, $cDDPM$, or $cGAN$; $condition$ indicates the input condition, i.e., region of interest ($ROI$) which controls the size of the generated scan and if it contains contrast or not, tumor segmentation ($seg$), or all previous conditions ($all$); $mode$ specifies how the conditioning is applied, such as wavelet transform ($w$), downsampling ($d$), convolutional layer ($conv$), or simple concatenation ($cat$); and $range$ represents the intensity clipping range, either 200 ([-200,200] HU) or 1000 ([-1000,1000] HU) for CT datasets, or simply $MRI$ for MRI datasets.

\subsubsection{Conditional Generative Adversarial Networks}
\label{subsec:cGAN}
A cGAN was trained to generate synthetic cases. The architecture of this GAN is based on the StyleGAN \citep{karras2019style}, but adapted to be conditional. The architecture of the cGAN is shown in \ref{app:cGAN_architecture}. 
For the training process, AdamW is used due to its stability, with batch size 1 and a learning rate of 0.0002, betas=(0.5, 0.999) \citep{ferreira2024gan}. Automatic mixed precision was not used as it leads to instability during training. For both datasets, 50 cases are randomly selected as the validation set, while the remaining cases are left for training. 
The loss function of the generator is formally defined by Equation \ref{eq:G_loss} and the discriminator by Equation \ref{eq:D_loss}.
\begin{equation}
\label{eq:G_loss}
\begin{split}
\mathcal{L}_{G} = & -\lambda_{1}\mathbb{E}{_{c,z}}[D(G(z|c))] \\
& + \lambda_{2}\mathbb{E}{_{x,c,z}}\left \| x-G(z|c)) \right \|_{MAE} \\
& + \lambda_{3}\mathbb{E}{_{x,c,z}}\left \| x-G(z|c)) \right \|_{tMSE}^2
\end{split}
\end{equation}
\begin{equation}
\label{eq:D_loss}
\begin{split}
\mathcal{L}_D = & -\lambda_{4}(\mathbb{E}{_{x,c}}[D(x|c)] - \mathbb{E}{_{z,c}}[D(G(z|c))]) \\
& + \lambda_{5} \mathbb{E}_{\tilde{x},c} [(||\nabla_{\tilde{x}} D(\tilde{x})||_2 - 1)^2] \\
& + 0.001*\mathbb{E}{_{x,c}}[D(x|c)]^2
\end{split}
\end{equation}
where  $D(x)$ is the output of the discriminator for a real ($x$) sample. $MAE$ is the mean absolute error, and $tMSE$ the mean squared error of the tumor.
$\tilde{x} \sim P_{\tilde{x}}$ are the samples uniformly drawn along straight lines between pairs of real and generated samples.
$\nabla_{\tilde{x}} D(\tilde{x})$ is the gradient of the discriminator's output with respect to the interpolated samples.
$||\nabla_{\tilde{x}} D(\tilde{x})||_2$ is the 2-norm (Euclidean norm) of the gradient. $0.001*\mathbb{E}{_{x,c}}[D(x|c)]^2$ is the regularization term to prevent the discriminator from outputting excessively large values. In our experiments, we employed different parameter settings for the CT and MRI cases to optimize the performance of our method. For the CT cases, the chosen parameters were: $\lambda_1 = 1$, $\lambda_2 = 1000$, $\lambda_3 = 100$, $\lambda_4 = 1$, and $\lambda_5 = 10$. For the MRI cases, the selected parameters were: $\lambda_1 = 1$, $\lambda_2 = 100$, $\lambda_3 = 100$, $\lambda_4 = 1$, and $\lambda_5 = 10$. We systematically tested multiple values for $\lambda_2$ and $\lambda_3$ — specifically, 1, 10, 100, and 1000 — to assess their impact on the results. The selected values were found to yield more realistic outcomes, particularly in the tumor region.

Iterative data augmentation was applied to the discriminator. Based on the works of \cite{karras2020training} and \cite{goceri2023medical}, non-leaky transformations are used, i.e., random rotation between [-6º, 6º], random shift between [-16, 16], random scaling of 0.2, random shear of 0.2 and random mirroring in all three directions. The probability of a transformation occurring is updated at each epoch, depending if overfitting is detected. In the event that the sum of the results from the training data is positive and the sum of the results from the validation data used is negative, the value probability increases $0.05$, in any other case it is decreased $0.05$. 
\subsubsection{Conditional denoising diffusion probabilistic model}
\label{subsec:cDDPM}
Two different approaches were tested with cDDPMs for the generation of synthetic cases with tumors. For the first approach, cDDPMs were developed based on 3D WDM \citep{friedrich2024wdm,friedrich2024cwdm}. Three approaches were therefore tested to concatenate the condition to the input: Apply the wavelet transform to the condition, creating a condition with the shape [24,128,128,128] ($WDM_{all\_w}^{200}$ and $WDM_{seg\_w}^{MRI}$); Down-sample (using nearest neighbors) the conditions to the same resolution, creating a condition of shape [3,128,128,128] ($WDM_{all\_d}^{200}$ and $WDM_{seg\_d}^{MRI}$ ); Use a 3D convolution (learnable layer) to reduce the shape of the conditions and out the condition with the shape [24,128,128,128] ($WDM_{all\_conv}^{200}$ and $WDM_{seg\_conv}^{MRI}$).
All conditions are only applied to the reverse process and concatenated to the input of the U-Net.

For the second approach, two cDDPMs are used. One model creates the synthetic scan where only the ROI and contrast are controlled, but not the tumor ($WDM_{ROI\_{d}}^{200}$ and $WDM_{ROI\_{d}}^{1000}$). The other model inserts synthetic tumors into the generated scans, i.e. it is a tumor inpainting model. In the last model, the wavelet transform is not used, instead random slices of shape $128^3$ are used to train the model, conditioned on the ROI, contrast, and the segmentation  ($DDPM_{all\_{cat}}^{200}$ and $DDPM_{all\_{cat}}^{1000}$). All cases with labels greater than $128^3$ were discarded for the inpainting model. The U-Net is similar in both cases, except for the number of layers, where the inpainting model has fewer one layer in the encoder and decoder. A random patch is selected in the full resolution scan for tumor placement. In \ref{app:inpaint} is presented the complete pipeline. Unrealistic regions, such as bone, are avoided by using the TotalSegmentator \citep{wasserthal2023totalsegmentator} to segment the main structures before selecting the location for the tumor inpainting.

The loss function used to train all models based on 3D WDM is formally defined by Equation 
\ref{eq:loss_c3DWDM_ct_generator}.
\begin{equation} 
\label{eq:loss_c3DWDM_ct_generator}
\mathcal{L}_{3D_{cWDM}} = \| \tilde{x}_0 - x_0 \|_2^2
\end{equation}
where $\tilde{x}_0 = \epsilon_\theta(x_t, t, c)$ represents the denoised scans, and $x_0$ denotes the real scans.

The loss function used to optimize the $DDPM_{all\_{cat}}$ models is formally defined by Equation \ref{eq:loss_c3DWDM_tumour_generator}.
\begin{equation} 
\label{eq:loss_c3DWDM_tumour_generator}
\mathcal{L}_{3D_{cDDPM}} = \| \tilde{x}_0 - x_0 \|_2^2 + \lambda_1 \|  \tilde{x}_0 \odot s - x_0 \odot s \|_2^2
\end{equation}
where $s$ is the segmentation mask. The loss function is the sum of the MSE between the real and denoised scan, and the MSE between the real and denoised tumor region, with a weighting factor of $\lambda_1=10$. The remaining hyperparameters are the same as defined by \cite{friedrich2024wdm}.

No data augmentation strategies were used to train the models with the MRI dataset.
However, conventional data augmentation strategies were used to train the models $WDM_{ROI\_{d}}^{200}$ and $WDM_{ROI\_{d}}^{1000}$. Due to the small amount of data and the high capacity of the U-Net, overfitting was considered. Therefore, a random ROI size was defined with scale\_range=((-0.05, 0.2), (-0.05, 0.2), (-0.05, 0.2)) and with a probability of 50\% for clipping or resizing with ‘nearest’.

Since random crops are selected from each scan for training the models $DDPM_{all\_{cat}}^{200}$ and $DDPM_{all\_{cat}}^{1000}$, random rotation in all axes, a 90-degree rotation, a 6-degree rotation in all axes, scale\_range=((-0.2,0.2),(-0.2,0.2),(-0.2,0.2)) and shear\_range=((-0.2,0.2),(-0.2,0.2),(-0.2,0.2)) were applied to the cropped scans with individual probability of 10\%.

The linear sampling method is used as baseline for all trained models with $T=1000$ steps.
DPM++ 2M \citep{lu2022dpm}, DPM++ 2M Karras (with Karras sigmas), DPM++ 2M SDE (with stochastic differential equation), and DPM++ 2M SDE Karras (which uses both) are also tested, with $T=100$.  

For tumor inpainting, patches with a size of $128^3$ were randomly selected from the full-resolution scan, with the label determining the shape of the tumor to be inpainted. To condition the generation of the synthetic tumor on the specific patch, the predicted region outside the tumor was replaced by the original volume at each step of the inference, as explained in \cite{lugmayr2022repaint}. However, an unrealistic change in intensities between the tumor and the surrounding tissue was observed. To avoid this sudden change, two strategies based on mask blurring were applied. To create the blurred mask, the original labeling was dilated by 5 iterations with a cubic structure of the form $3^3$ and a Gaussian blur with a blur factor of 25 was applied to the mask. The ‘edge blur’ ensures that all areas into which the tumor is to be inserted have the value 1, but they are not taken into account in the ‘full blur’. The final mask has values ranging from 0 to 1, where 1 indicates that the value is completely replaced by noise at time step $t$ during inference, and 0 indicates that the value is completely replaced by the original value.

\subsection{Radiomics features Test}
\label{subsec:Radiomic_feature_test}
Features extracted from all real and synthetic datasets were used to evaluate the quality of the synthetic data. For feature extraction, the scans were discretized using a fixed bin width of 25 and analyzed within a binary mask (label = 1) with mask correction enabled. Features were computed on the original images, Laplacian-of-Gaussian–filtered images ($\sigma=1–5 mm$), and full wavelet decompositions. Extracted feature families included 3D shape descriptors, first-order intensity statistics, 22 Gray Level Co-occurrence Matrix metrics (e.g., Autocorrelation, Contrast, Cluster Prominence, Joint Entropy, Informational Measure of Correlation 1 and 2, and Sum Squares), and all non-redundant features from the Gray Level Run Length Matrix, Gray Level Dependence Matrix, and Neighborhood Gray Tone Difference Matrix classes. A total of 1,065 radiomic features were extracted per ROI under the custom settings (binWidth = 25; label = 1; correctMask = True). 

Principal Component Analysis (PCA) and CCC are used to visually capture the differences between the extracted features. It is evaluated whether the performance in the segmentation task correlates with the CCC between the real and synthetic features. Moreover, if the synthetic data and the real data exhibit similar distributions, the generated synthetic data can also be effectively utilized downstream tasks using Radiomics.

\subsection{Segmentation Test}
\label{subsec:seg_test}
\subsubsection{Creation of ground truth}
The real datasets $CT_{real}^{778}$ and $MRI_{\text{real}}^{1000}$ contain the respective tumor labels that are used to train the segmentation models for tumor segmentation of both the real and synthetic cases, as these labels are used as condition of the generative models.

The real ($CT_{real}^{1063}$) and synthetic CT datasets generated by the cDDPMs are also used to train the segmentation model for the segmentation of bone structures. Three tools are used to generate the ground truth, namely TotalSegmentator \citep{wasserthal2023totalsegmentator}, TotalSpineSeg \citep{Warszawer2024} and AMASSS-CBCT \citep{gillot2022automatic}. The structures tested with the individual tools are listed in the supplementary table \ref{tab:DSC_bone_exp}. 

\subsubsection{nnU-Net training}
To test the applicability of the synthetic scans in real tasks, segmentation models are trained for the two different modalities. 
The datasets generated with cGANs and cDDPMs are used to train a full-resolution nnU-Net 3D network for tumor and bone segmentation \citep{isensee2021nnu}. All trained models and the respective tasks are listed in supplementary table \ref{tab:exp_summary}. It is tested whether the DSC of a model trained with synthetic data shows similar results to those of a model trained with real data. The real test data is used to calculate the DSC. The datasets were adapted to the format of nnU-Net and the remaining training steps were decided automatically by nnU-Net. 

\subsection{Visual Turing Test}

The VTT was performed by five experienced radiologists (C.W., F.O., T.O., M.B., and R.S.), consisting of two subtasks: Image classification, where 100 real and 100 synthetic random slices (from random axes) were presented to the experts who had to rate them on a scale from 1 to 10; Scan Classification, where 100 real and 100 synthetic random scans were provided to the experts, which they could load into any viewer of their choice and rate each scan on the same 1 to 10 scale. The 1 indicates a completely realistic/real image, and 10 indicates a completely unrealistic/synthetic image. 
Values bellow or equal to 5 are considered as “Real”, while values above 5 are considered as “Synthetic”. The number of real and synthetic cases was not disclosed to the experts.

\subsection{Evaluation Metrics }
\label{subsec:evaluation_metrics}
In addition to DSC, PCA and CCC, the synthetic cases are evaluated using MAE and MS-SSIM. These metrics are employed to assess the similarity between synthetic and real cases, as well as to evaluate whether the distribution of synthetic datasets aligns with that of real datasets.

\subsection{Workstation}
A cluster node with 6 NVIDIA RTX 6000, 48 GB of VRAM, 1024 GB of RAM, and AMD EPYC 7402 24-Core Processor and a cluster node equipped with NVIDIA H100 GPUs with 96 GB of VRAM, 512 GB of RAM and CPUs Intel Xeon 8468 Sapphire were used. Two GPUs were required for each experiment with GANs, one for the generator and the other for the discriminator. Only one GPU was required for the DDPMs and the nnU-Net training. Python 3.10.13, MONAI 1.3.0, PyTorch 2.5.1, nnUNetv2 and PyRadiomics 3.0.1 were used in the experiments.

\section{Results}
\label{sec:results}
\subsection{Radiomic features between real and synthetic data}

\begin{figure}[h]
\centering
\includegraphics[width=0.48\textwidth]{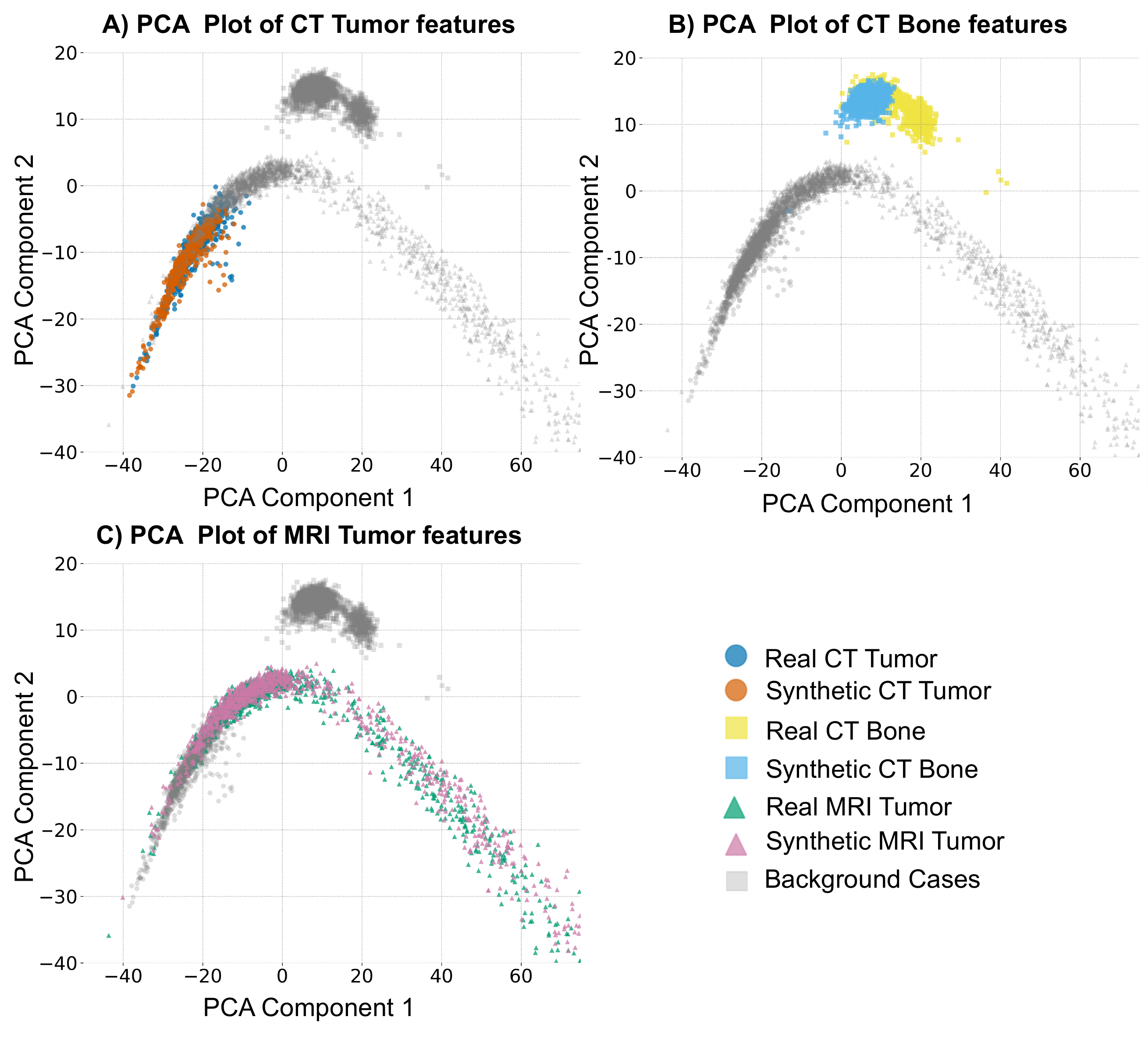}
\caption{PCA plots of Radiomic features. Each plot compares each pair of real and synthetic feature distributions. Plot A) presents the features of the tumor region and B) the bone region of CT scans, and C) the tumor region of MRI scans.}\label{fig:pca_plot}
\end{figure}

The synthetic datasets were first evaluated by visually inspecting the generated cases followed by Radiomic features mining. We found that the generative networks were better able to capture and reproduce the Radiomic features of MRI scans rather than CT scans. Fig. \ref{fig:pca_plot} presents the PCA of Radiomic features extracted from A) tumor region and B) bone region of CT scans, and C) tumor region of MRI scans. Plot C presents a greater overlap and smaller Euclidean distance between real and synthetic pairs than the remaining plots ($C=1.08 < A=4.06 < B=6.79$). Considering only the regions to study in the segmentation task, it can be seen that features of synthetic and real tumors present a greater overlap than bone, specially for MRI. 

\begin{figure*}[t]
\centering
\includegraphics[width=0.9\textwidth]{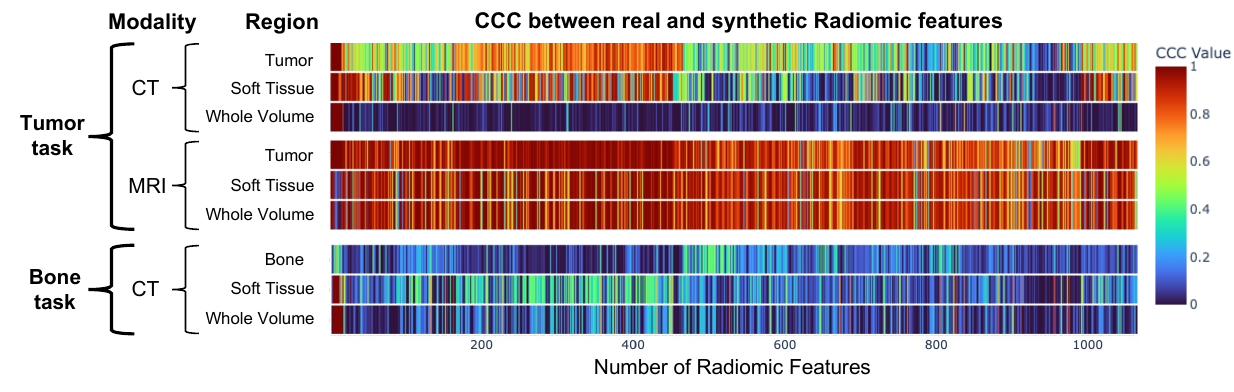}
\caption{Concordance Correlation Coefficient (CCC) analysis of Radiomic features extracted from CT and MRI scans across different regions. The heatmap visualizes the correlation results, with stronger correlations indicated in red and weaker correlations in black. Only the datasets with best performance on each task are presented.}
\label{fig:ccc}
\end{figure*}

The CCC results were categorized as excellent (CCC~$\geq$~0.9), Good (0.75~$\leq$~CCC~$<$~0.9), Moderate (0.5~$\leq$~CCC~$<$~0.75), and Poor (CCC~$<$~0.5). As illustrated in Fig. \ref{fig:ccc}, synthetic MRI features exhibit a higher degree of correlation, specially in the tumor region with 668 categorized as excellent, 242 as good, 123 as  
moderate and only 32 as poor. In contrast, the synthetic CT dataset with best performance on the tumor segmentation task presents only 98 excellent, 167 good, 352 moderate and 448 poor features. Similarly, the best synthetic dataset in the bone segmentation task also presents a high number of poor features (1063) with only 2 categorized as moderated and none as good or excellent. This emphasizes the low levels of detail of synthetic CT scans compared to synthetic MRI scans.

The synthetic CT datasets present a MAE below 0.4 HU and a MS-SSIM close to the real value (0.3581), which would indicate a similar distribution of the synthetic datasets to the real dataset. Similarly, the MRI scans have MAE between 0.02 and 0.08 and MS-SSIM values also close to the real dataset (0.7201). However, this metrics does not ensure utility. These results are summarized in \ref{app:results}.

\subsection{Segmentation Test}
\begin{figure}[h]
\centering
\includegraphics[width=0.48\textwidth]{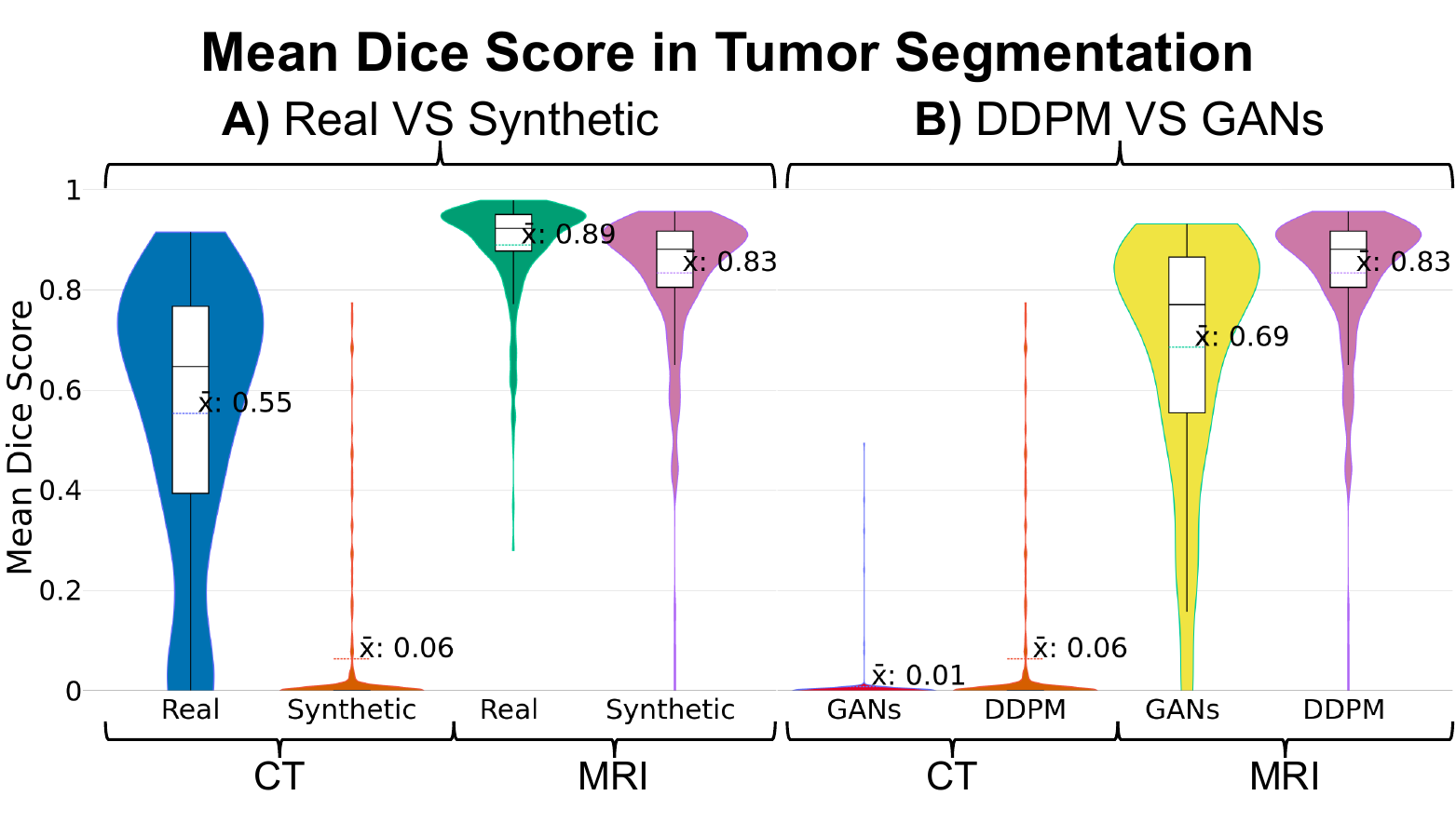}
\caption{Comparison of Dice scores for tumor segmentation on the test set. A) Comparison of models trained on real data versus the best-performing models trained on synthetic data. B) Comparison of the best-performing models trained on synthetic data generated by GANs versus generated by DDPMs.}\label{fig:dsc_tumor}
\end{figure}

To assess the usefulness of synthetic data, segmentation tests were performed. We found a association between the CCC of Radiomic features and DSC for the segmentation of tumors. CT scans have a poor CCC leading to low performance in the segmentation task (DSC=0.06), in comparison with the real dataset (DSC=0.55). MRI have excellent CCC leading to high performance (DSC=0.83), comparable to the real dataset (DSC=0.89) as illustrated in Fig. \ref{fig:dsc_tumor} A). Moreover, Fig. \ref{fig:dsc_tumor} B) shows that data generated by DDPMs produces better DSC than data generated by GANs.

\begin{figure}[h]
\centering
\includegraphics[width=0.48\textwidth]{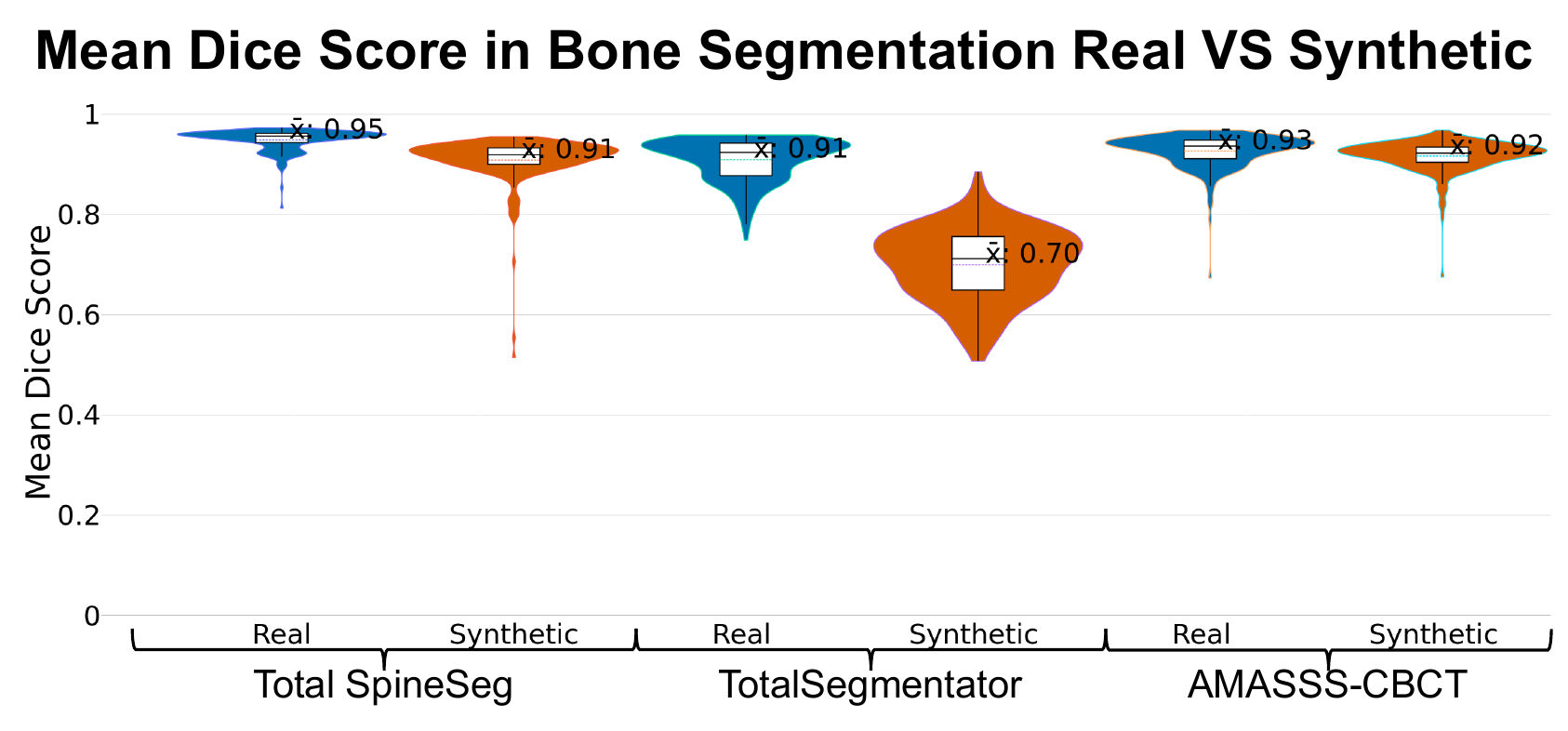}
\caption{Comparison of Dice scores for bone structure segmentation on the test set using models trained on real data (blue) and the best-performing model trained on synthetic data (orange). Three tools were used for the creation of the ground truth: Total SpineSeg, TotalSegmentator and AMASSS-CBCT.}\label{fig:dsc_bone}
\end{figure}

The same association is not entirely valid for bone segmentation. Although the CT scans have poor CCC, the performance of the segmentation models trained on synthetic cases are comparable with the models trained on real data, as can be seen in Fig. \ref{fig:dsc_bone}. This is related with the complexity of the task. A substantial discrepancy in DSC values is observed when using ground truth from the "TotalSegmentator" tool: synthetic data attain a DSC of 0.7, markedly lower than the 0.91 DSC achieved by the real dataset. These labels are inherently more complex than those produced by the "Total SpineSeg" and "AMASSS-CBCT" tools. These achieved DSCs of 0.91 and 0.92 on synthetic data, closely approximating the real dataset’s DSCs of 0.95 and 0.93, respectively.

\begin{figure}[h!]
\centering
\includegraphics[width=0.48\textwidth]{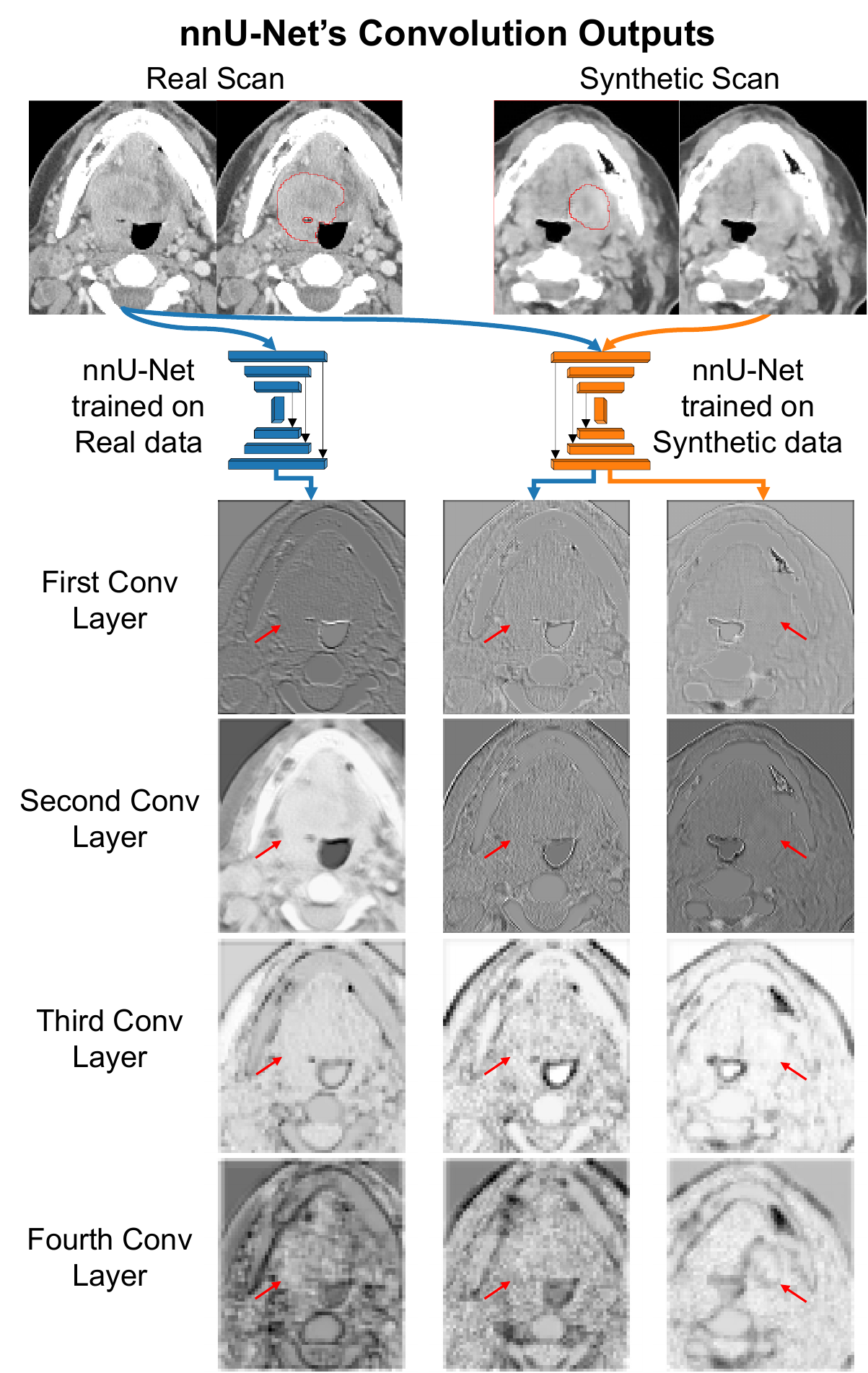}
\caption{Mean feature maps extracted from the first to fourth convolutional layers of trained nnU-Net models. The first column displays features from a real CT scan processed by a model trained on real data. The second column shows features from the same real scan processed by a model trained on synthetic data. The third column presents features from a synthetic CT scan processed by a model also trained on synthetic data. All models were trained for tumor segmentation in CT images. Tumor regions are highlighted with red borders, and red arrows indicate the tumor location.}\label{fig:nnunet_features}
\end{figure}

Defined borders, i.e. high-contrast of the region to segment, are very important for segmentation models. It was observed that the tumors in synthetic CT scans exhibited too well defined borders across all techniques employed for synthetic dataset creation. In contrast, real cases do not exhibit such well-defined borders. This discrepancy resulted in an easier training task but a significantly more challenging testing task. Upon analyzing the initial layers of the segmentation models trained for tumor segmentation on CT scans using synthetic data, it was evident that these models placed a high emphasis on the unrealistic, well-defined borders present in the synthetic scans. These boundaries are not present in real CT scans, which causes the models to remove the tumor in the first layers of the encoder. Consequently, this made it impossible for the segmentation models to perform well on real cases. This phenomenon can be visualized in Fig. \ref{fig:nnunet_features}. The same is true for bone structures, but both real and synthetic bones have well defined HU values and borders.

\subsection{Visual Turing Test evaluation}
\begin{figure}[h]
\centering
\includegraphics[width=0.48\textwidth]{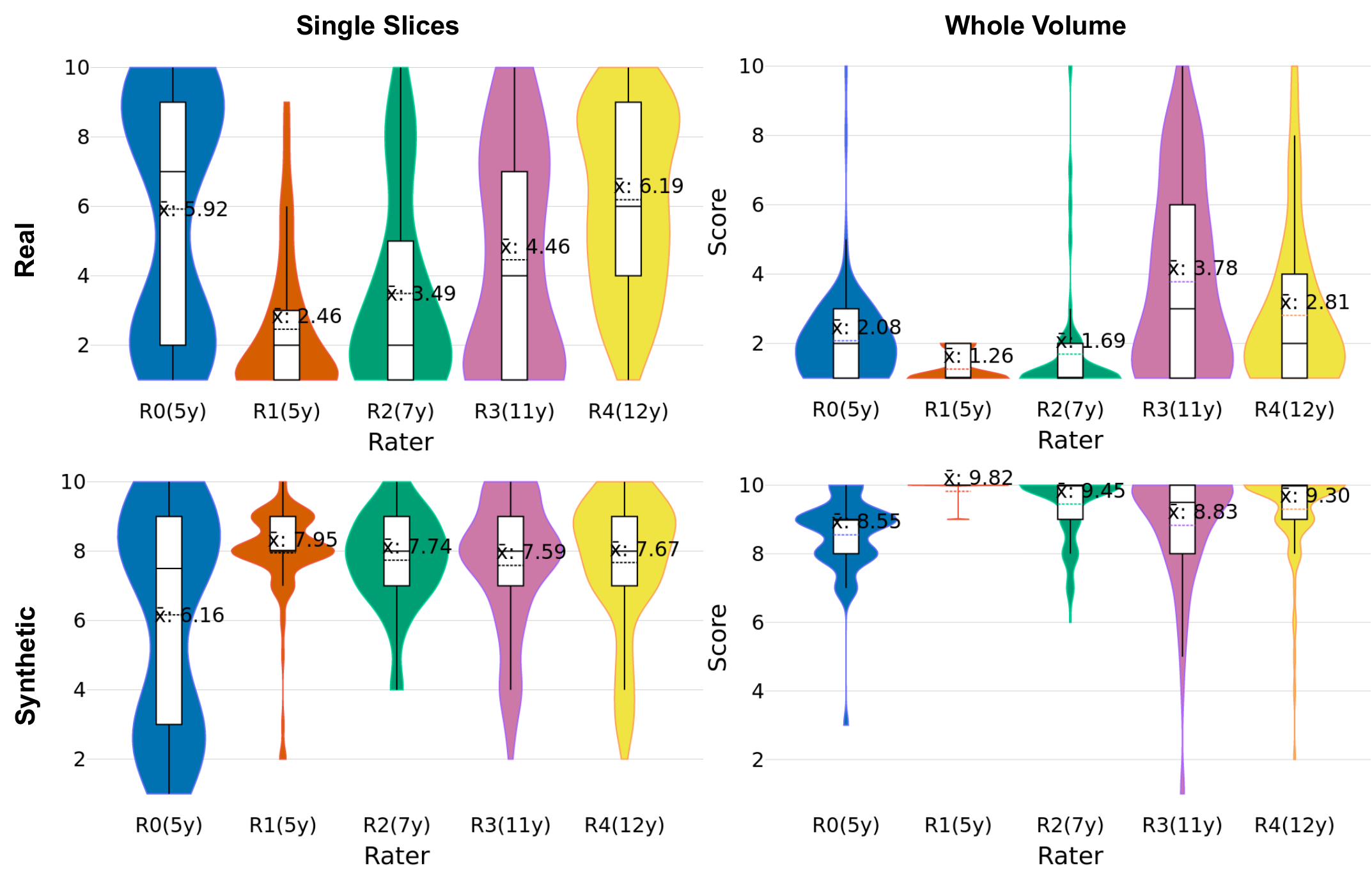}
\caption{Visual Turing Test performance by each rater (R), ordered by total years of practice and synthetic-data experience.
Top row: classification of real cases; bottom row: synthetic cases.
Left column: per-slice evaluations; right column: whole-volume evaluations. 1 indicates a completely realistic/real image, and 10 indicates a completely unrealistic/synthetic image.}\label{fig:vtt}
\end{figure}

Due to the limited results of CT scans in tumor segmentation as well as the poor correlation of Radiomic features, the synthetic full-resolution CT scans generated by diffusion models with the best performance in tumor segmentation were subjected to VTT to determine whether they were realistic enough to be considered real cases for educational purposes. Fig. \ref{fig:vtt} illustrates that expert raters discriminated synthetic from real images with markedly greater accuracy when provided with full volumetric data than with single slices. On binary classification, inter-rater agreement increased from Fleiss's K = 0.303 at the slice level to K = 0.778 for whole-volume assessments, underscoring the importance of volumetric context for reliable evaluation. 

Statistical significance was evaluated using independent-samples t-tests and Mann–Whitney U tests. The difference in ratings between real and synthetic cases was highly significant ($p<0.001$). A similarly strong effect was observed when comparing raters with more than ten years of experience to those with less ($p<0.001$). However, applying a more stringent significance threshold of $\alpha$$=0.04$, no significant difference remained between the two experience groups for binary whole-volume classifications.

\section{Discussion}
\label{sec:discussion}
We show the importance of assessing the quality of synthetically generated data using distinct approaches, as using a limited range of tests can lead to misleading conclusions. Synthetic data generated by GANs and DDPMs have great potential to replace real data for real uses, however still limited by the complexity of the task. Although the synthetic data looked realistic, deeper analysis shown lack of detailed reproduction of essential features to ensure their utility. 

The extraction and analysis of Radiomic features shown the high difficulty of the generative networks to reproduce realistic features of CT scans. Fig. \ref{fig:pca_plot} show how easily synthetic CT cases can be separated from real ones considering the bone features. The correlation seen in Fig. \ref{fig:ccc} also supports the lack of reproduction of features in synthetic CT cases, for both tumor and bone segmentation tasks. Such results were expected because of the higher complexity of the CT dataset in comparison with the MRI dataset. The distribution of the datasets can be measured by the MS-SSIM metric. The MRI dataset has less variability (MS-SSIM=0.7201), while the CT dataset has highly variable cases (MS-SSIM=0.3581), making it more difficult for generative networks to capture all important features in CT scans.

The segmentation of tumors on CTs was very challenging even when using real data, as shown by the real DSC of 0.55. The high complexity of such task demands high fidelity of the synthetic data, which was not enough to allow nnU-Net to generalize to real samples. Both DDPMs and GANs architectures failed to produce satisfactory results. Nevertheless, it was found that DDPM without wavelet transform achieves the best DSC=0.06, which indicates that using the wavelet transform to generate the synthetic cases introduces unrealistic features. The unrealistic features created by the wavelet based DDPMs are even detected visually. As shown in Fig. \ref{fig:nnunet_features}, as well in the \ref{app:Synthetic_CT_scans_tumour}, the synthetic cases generated with wavelet transform exhibit borders over defined. These unrealistic feature is less accentuated in the DDPM without wavelet transform, however, still too accentuated. As observable in the region highlighted by the red circle and in subsequent images in Fig. \ref{fig:nnunet_features}, the segmentation network captures and learns to accentuate the unrealistic contrast of the tumor, as seen in the right column. However, when tested on real data lacking this characteristic, the segmentation model fails, progressively removing the tumor at each layer of the U-Net, as observed in the middle column. Upon closer inspection of the left column, it is evident that the model trained on real data learns to enhance the contrast of the tumor region. Despite achieving a MAE below 0.4 HU and a MS-SSIM close to the real value (0.3581), indicating that synthetic datasets are highly similar and exhibit comparable distributions, these metrics do not necessarily signify sufficient realism for downstream tasks. These broadly used quantitative metrics are not capable of ensuring usability of the synthetic data, in contrast with the Radiomic features CCC.

It was noticed that the features with worst correlation were associated with texture, e.g., the Grey Level Co-occurrence Matrix Imc1, Imc2 and the inverse variance extracted with the wavelet filter. The kurtosis feature also show that the synthetic cases have smoother textures with less sharp intensity changes, i.e. less pronounced details, possibly indicating a lack of realistic noise, fine structure or micro-texture details. This is a typical feature of synthetic images where the models aim to produce visually appealing but less natural looking textures. These properties apply to both GANs and diffusion models.

The segmentation of brain tumors on MRI scans using synthetic cases yielded competitive results compared to using real cases. This task is less challenging than the segmentation of tumors on H\&N CT scans, as evidenced by the high DSC of 0.89 achieved by the model trained on real data, as well as the MS-SSIM value of 0.7201, which indicates that the MRI dataset is less variable, making it easier for the generative models to learn the real distribution.  Nevertheless, the high DSC achieved by the best model trained on synthetic cases suggests that the synthetic cases exhibit higher fidelity to the real cases, which is supported by the CCC shown in Fig. \ref{fig:ccc}. The segmentation models trained on synthetic data generated by GANs yielded worse results compared to those trained on synthetic data produced by DDPMs. The MS-SSIM of 0.8551 shows that GANs were not able to reproduce the variability of the original MRI dataset.
The WT, composed by the ET, NCR and ED, shows the greatest difference between real and synthetic data. This was expectable as the ED contains a hyperintense signal in the FLAIR modality and is more difficult to recognize on the used T1c scans. With a lower contrast difference between the ED and the surrounding tissue, the generative networks had greater difficulty generating this region with higher fidelity, leading to poorer segmentation results. Some synthetic examples can be seen in \ref{app:Synthetic_MRI_scans}.
It is also important to notice that the sampling methods ‘DPM++ 2M’ and ‘DPM++ 2M SDE’ show consistently better results than the other methods for all three models.
A narrower distribution of cases and a higher similarity between real and synthetic MRI scans recognizable by the high MS-SSIM and low MAE, respectively,  are among the main reasons why the generative models were able to better learn the distribution of the MRI data.

The segmentation tests on bone structures shows the possibility of using synthetic data for this task. The ground truth labels created with the "TotalSegmentator" for the test cases were verified by two radiologists to ensure that a fair and realistic comparison could be made between models. The vertebrae, spinal canal, and clavicula had perfect segmentations, while the skull bases and sternum had some missing parts or incomplete segmentations, and the costosternal joints were often incomplete or over-segmented, with a few cases of contrast agents detected as bone. 
Fig. \ref{fig:dsc_bone} shows comparable performance of the segmentation model trained on real and synthetic data for the ground truth structures created with the "Total SpineSeg" and the "AMASSS-CBCT" models. However, a discrepancy of DSC=0.21 is observable for the structures segmented with the "TotalSegmentator". It was observed that the segmentation models trained with ground truth labels generated using the "TotalSegmentator" encountered difficulties in differentiating between the left and right sides for the ribs and clavicula. Additionally, in certain instances, multiple ribs were segmented under the same label. These issues emphasize the importance of high correlation, even for bone features. Figure \ref{fig:ccc} illustrates poor CCC for the bone. Although easier tasks such as spine or jawbone segmentation achieved high DSC, more complex segmentation tasks (including the clavicle, and ribs) resulted in lower DSC scores. Bones are characterized by high HU values compared to other tissues, as well as high contrast, which facilitates segmentation. However, these characteristics result in limited performance when the complexity of the task increases and the remaining features were not reproduced.

The analysis of the extracted features show high differences of the wavelet-based total energy feature which suggests that synthetic images can alter density distribution profiles, which is a known limitation of current synthesis techniques that prioritize global intensity adjustment over local energy variations \citep{yuan2023comprehensive}.
Synthetic cases present more uniform texture patterns, as shown by the lower ‘RunLengthNonUniformity’ values, possibly oversimplifying the natural texture complexity. 

The lack of realistic features and poor performance on the segmentation task suggest limited used of synthetic data for education purposes, which is also supported by the VTT. Raters with more experience were more likely to classify real cases as synthetic, reflecting the unrealism of some real cases. However, no significant difference between more as less experienced raters was found for classification between real and synthetic cases, with substantial agreement in scan classification. It was highlighted by the experts the low contrast and low vessel visibility of the synthetic cases, which helped to distinguish each case. Some anatomic mismatch was also used as indicator of synthetic cases. Artifacts created by beam hardening were also indicator of realism, as real cases had a sharper delineation of these artifacts. 
The nature of the DDPMs' loss function leads to smooth/blurred generation, even though realistic, it can be detected by trained experts. The lack of realism in the tumor border would unrealistically bias the students to find these on real cases, making the training inefficient, and synthetic data not useful. Some examples can be seen in \ref{app:Synthetic_CT_scans_bone}.

The better performance of bone segmentation compared to tumor segmentation using synthetic CT scans can be attributed to the higher complexity of the CT in comparison with the MRI dataset, the low correlation of Radiomic features on the synthetic CT cases, as well as the inherent difficulty of the task at hand. 
The segmentation networks rely on the higher contrast of the region to segment. However, not all structures have this features, which is the case of tumor on CT.
Bone segmentation is generally easier for DL networks as the contrast is higher and the boundaries of the bones are clearer compared to the surrounding tissue, in both real and synthetic cases. In contrast, tumor segmentation is more difficult as tumor tissue often has similar HU values to the surrounding tissue and less clear boundaries. 
Seen this, we conclude that synthetic data generated by DDPMs can be created locally and shared with ensured anonymity and utility, although limited to specific downstream tasks.

\section{Conclusion}
In this study, we demonstrate that synthetic data can be used independently for the segmentation task, although limited by the complexity of the structures to segment. 
The generative models were better able to capture the Radiomics features of MRI scans due to the lower heterogeneity of the dataset in comparison with the CT dataset which presented poor correlation with the real cases. This led to strong segmentation performance for tumors on MRIs, but poor performance on CTs, with DDPM consistently outperforming GAN-based approaches overall. On the other hand, the segmentation of bone structures using synthetic CT scans presented competitive results to the use of real data. This highlights that the more complex the task, the more accurately the features must be reproduced. 
The VTT also supports the limited use of synthetic CT scans, which, although visually realistic, were noted by radiologists to exhibit low soft-tissue contrast. This reflects a common limitation of current state-of-the-art generative models, which tend to prioritize visual appeal over the accurate reproduction of natural tissue textures. 
Therefore, in the future we need generative models that are able to learn more heterogeneous datasets and not only focus on the overall aspect of the scan.

\section*{Acknowledgements}

André Ferreira was supported by FCT - Fundação para a Ciência e Tecnologia, I.P. by project reference 2022.11928.BD and DOI identifier https://doi.org/10.54499/2022.11928.BD.
Behrus Puladi was funded by the Medical Faculty of RWTH Aachen University as part of the Clinician Scientist Program. 
This work has been supported by FCT – Fundação para a Ciência e Tecnologia within the R\&D Unit Project Scope UID/00319/Centro ALGORITMI (ALGORITMI/UM).
It was also supported by the Advanced Research
Opportunities Program (AROP) of RWTH Aachen University.

\section*{Code availability }
The code used in this work is available in an open-source GitHub
repository at \href{https://github.com/ShadowTwin41/generative_networks}{https://github.com/ShadowTwin41/generative\_networks}.

\section*{Ethics approval} This research did not involve human participants, animal subjects, or sensitive data, and therefore did not require ethical approval. All data used in this study are publicly available.
\\
\section*{Competing interests} The authors declare no competing interests.

\section*{Glossary}
\begin{itemize}
    \setlength{\itemsep}{-3pt}   
    \item AI - Artificial Intelligence
    \item BraTS - Brain Tumor Segmentation
    \item CCC - Concordance Correlation Coefficient
    \item cDDPMs - conditional Denoising Diffusion Probabilistic Models
    \item cGANs - conditional Generative Adversarial Networks
    \item CT - Computed Tomography
    \item DDPM - Denoising Diffusion Probabilistic Models
    \item DL - Deep Learning
    \item ED - Peritumoral Edema
    \item ET - Enhancing Tumor
    \item FL - Federated Learning
    \item GANs - Generative Adversarial Networks
    \item HU - Hounsfield Units
    \item MAE - Mean Absolute Error
    \item ML - Machine Learning
    \item MRI - Magnetic Resonance Imaging
    \item MS-SSIM - Multi-Scale Structural Similarity Index Measure
    \item NCR - Necrotic Tumor Core
    \item PCA - Principal Component Analysis
    \item PSNR - Peak Signal to-Noise Ratio
    \item RMSE - Root Mean Squared Error
    \item ROI - Region of Interest
    \item SSIM - Structural Similarity Index
    \item TC - Tumor Core
    \item TCIA - The Cancer Imaging Archive
    \item VTT - Visual Turing Test
    \item WT - Whole Tumor
\end{itemize}

\section*{Author contributions: CRediT}
\begin{itemize}
    \setlength{\itemsep}{-3pt}   
    \item \textbf{Conceptualization} – A.F., B.P., V.A., J.E.
    \item \textbf{Data curation} – A.F., K.X.,  C.W., G.C.
    \item \textbf{Formal analysis} – A.F., B.P., V.A., J.E.
    \item \textbf{Funding acquisition} – B.P., V.A., J.E.
    \item \textbf{Investigation} – A.F., C.W., F.O, T.O, M.B., R.S.
    \item \textbf{Methodology} – A.F., B.P., K.X., V.A.
    \item \textbf{Project administration} – B.P.
    \item \textbf{Resources} – B.P., V.A., J.E., F.H., R.R., J.K.
    \item \textbf{Software} – A.F.
    \item \textbf{Supervision} – B.P., V.A., J.E.
    \item \textbf{Validation} – A.F., B.P., C.W., G.C.
    \item \textbf{Visualization} – A.F., B.P.
    \item \textbf{Writing – original draft} – A.F.
    \item \textbf{Writing – review and editing}– B.P, A.F., K.X., C.W., G.C., F.O., T.O., M.B., R.S., F.H., R.R., J.K, D.T, J.E., V.A.
\end{itemize}

 \bibliographystyle{elsarticle-num-names.bst}\biboptions{authoryear}
\bibliography{bib}

\appendix
\onecolumn
\section{Results tables}
\label{app:results}
The downstream tasks and respective models used to generate the synthetic dataset are summarized in Table \ref{tab:exp_summary}. The bone structures are specified in Table \ref{tab:DSC_bone_exp}. 
The Tables \ref{tab:ct_seg_and_metrics} and \ref{tab:mri_seg_results} show the DSC for tumor segmentation on the $CT_{real}^{195}$ and $MRI_{real}^{{251}}$, respectively. The MAE and MS-SSIM metrics are also presented. These experiments use the label as condition for the generation of the synthetic tumor.

The Table \ref{tab:inpainted_dsc} shows the DSC for the inpainting models. The first two synthetic datasets were created with the model $WDM_{ROI\_d}^{200}$ (or $WDM_{ROI\_d}^{1000}$) for creating the full-resolution scan without the tumor and $DDPM_{all\_{cat}}^{200}$ (or $DDPM_{all\_{cat}}^{1000}$) for inpainting the synthetic tumor. The last two datasets were only generated with $DDPM_{all\_{cat}}^{200}$ and $DDPM_{all\_{cat}}^{1000}$. These datasets are not the full resolution, but only patches of shape $128^3$. For the training of the nnUNet, each scan was considered a full patch.

The Table \ref{tab:ct_metrics} presents the MAE and MS-SSIM metrics for the CT datasets used for the bone segmentation task. Tables \ref{tab:bone_seg_results} and \ref{tab:bone_seg_results_2} show the DSC for the segmentation of bone structures on the $CT_{real}^{195}$.

\begin{figure}[h]
\centering
\includegraphics[width=0.8\linewidth]{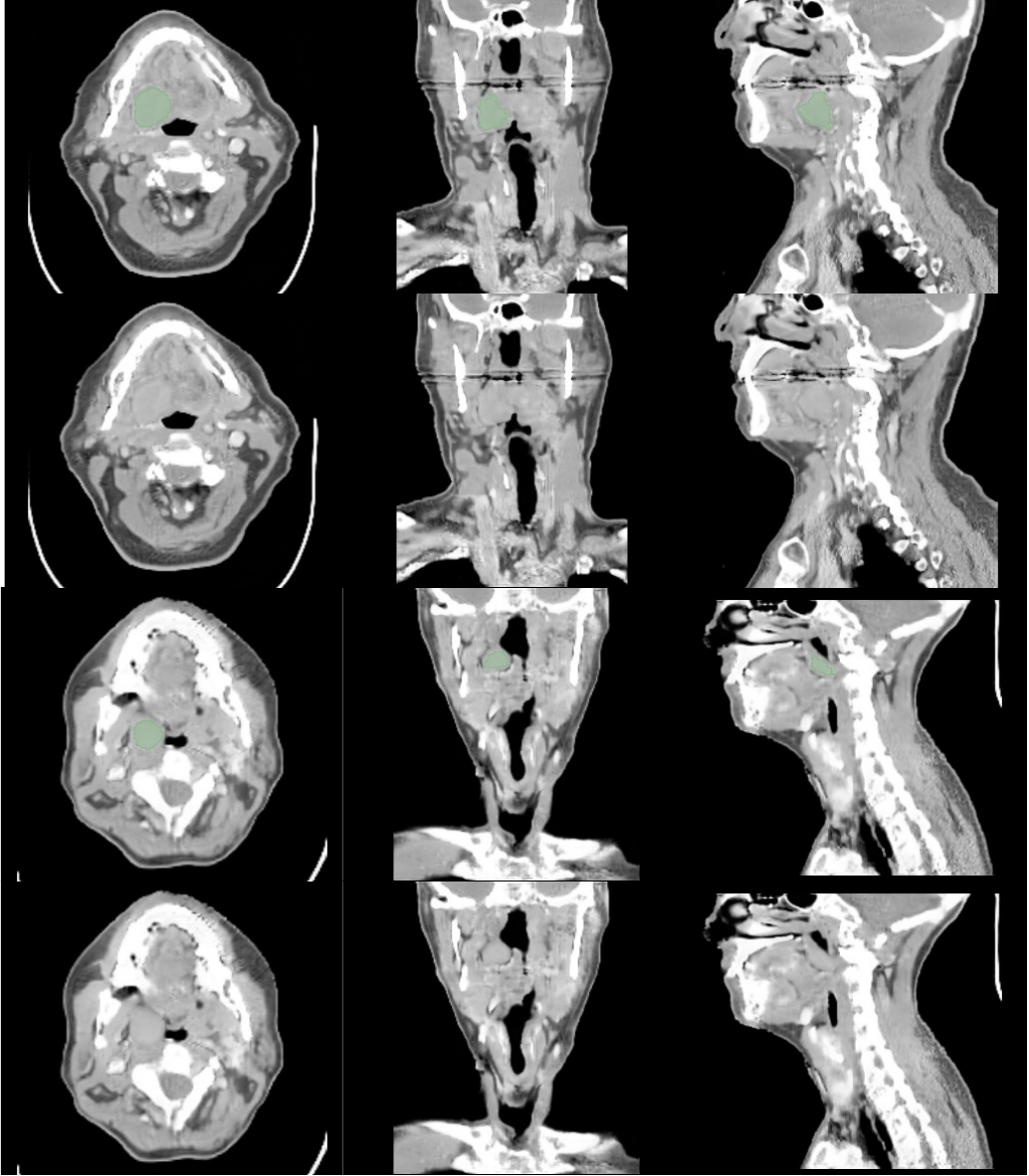}
\caption{Slices in the three axes of the synthetic CT scans generated using $cGAN_{seg}^{MRI}$. The tumor is segmented in green.}
\label{fig:app:fake_tumour_GAN}
\end{figure}

\begin{figure}[h]
\centering
\includegraphics[width=0.8\linewidth]{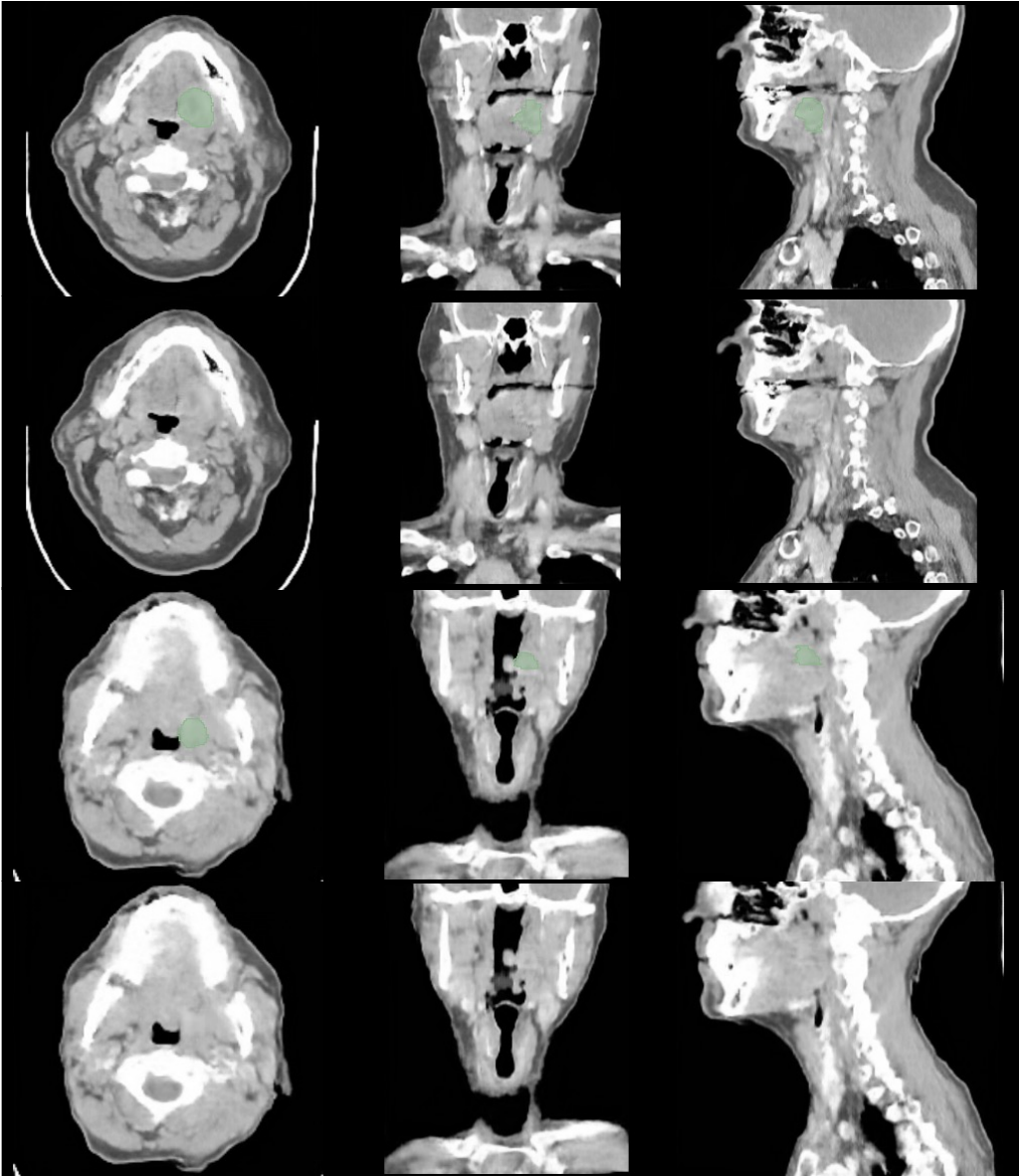}
\caption{Slices in the three axes of the synthetic CT scans generated using $WDM_{all\_conv}^{200}$ and the sampling method DPM++ 2M. The tumor is segmented in green.}
\label{fig:app:fake_tumour_convcat}
\end{figure}

\begin{figure}[h]
\centering
\includegraphics[width=0.8\linewidth]{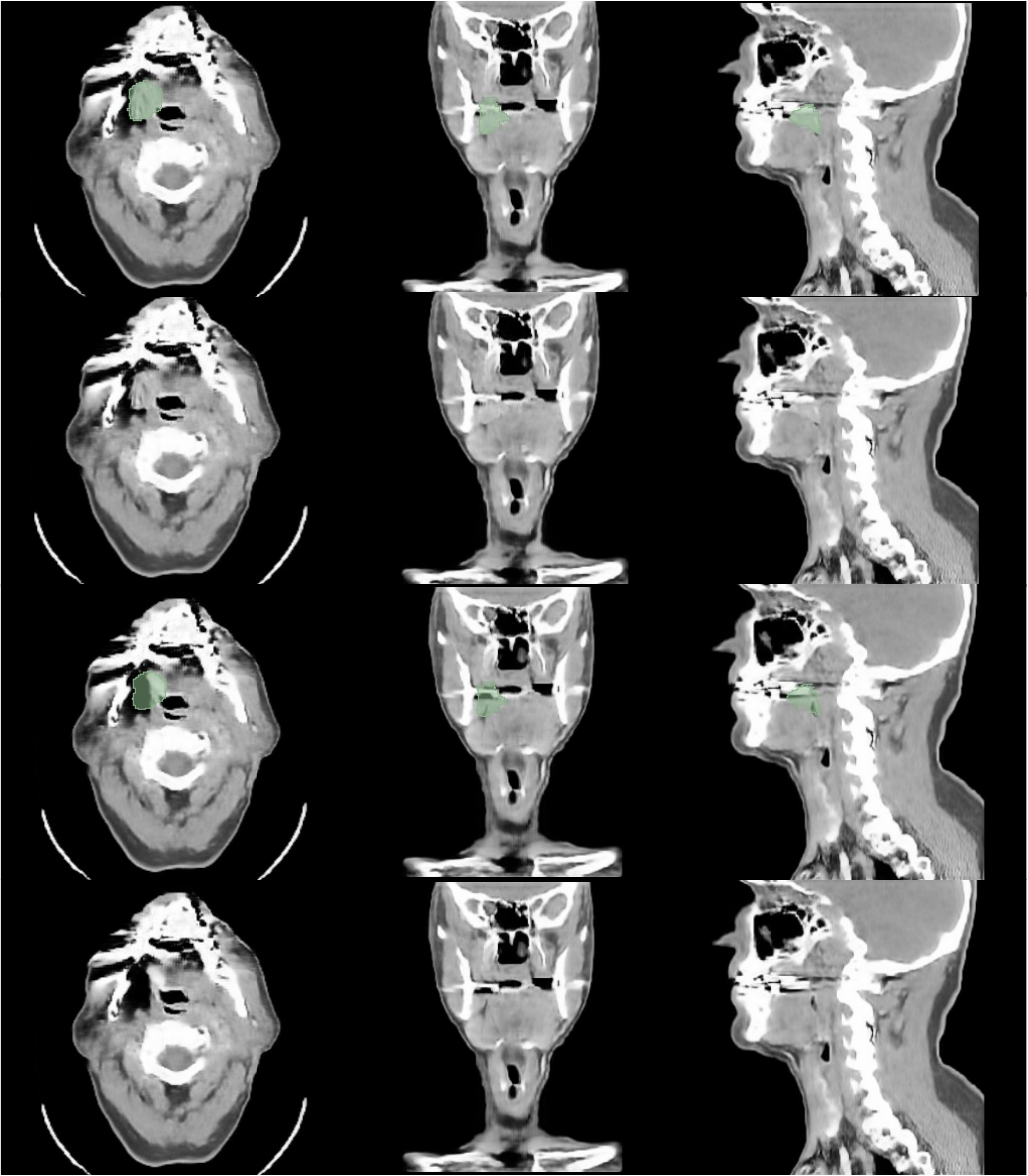}
\caption{Slices in the three axes of the synthetic CT scans generated using $WDM_{ROI\_d}^{200}$ to generate the full resolution scans and  $DDPM_{all\_{cat}}^{200}$ for inpainting, both using the sampling method Linear. The ‘Edge blur’ mask at the top and the ‘Full blur’ mask at the bottom.  The tumor is segmented in green.}
\label{fig:app:fake_tumour_inpainted}
\end{figure}

\begin{figure}[h]
\centering
\includegraphics[width=0.8\linewidth]{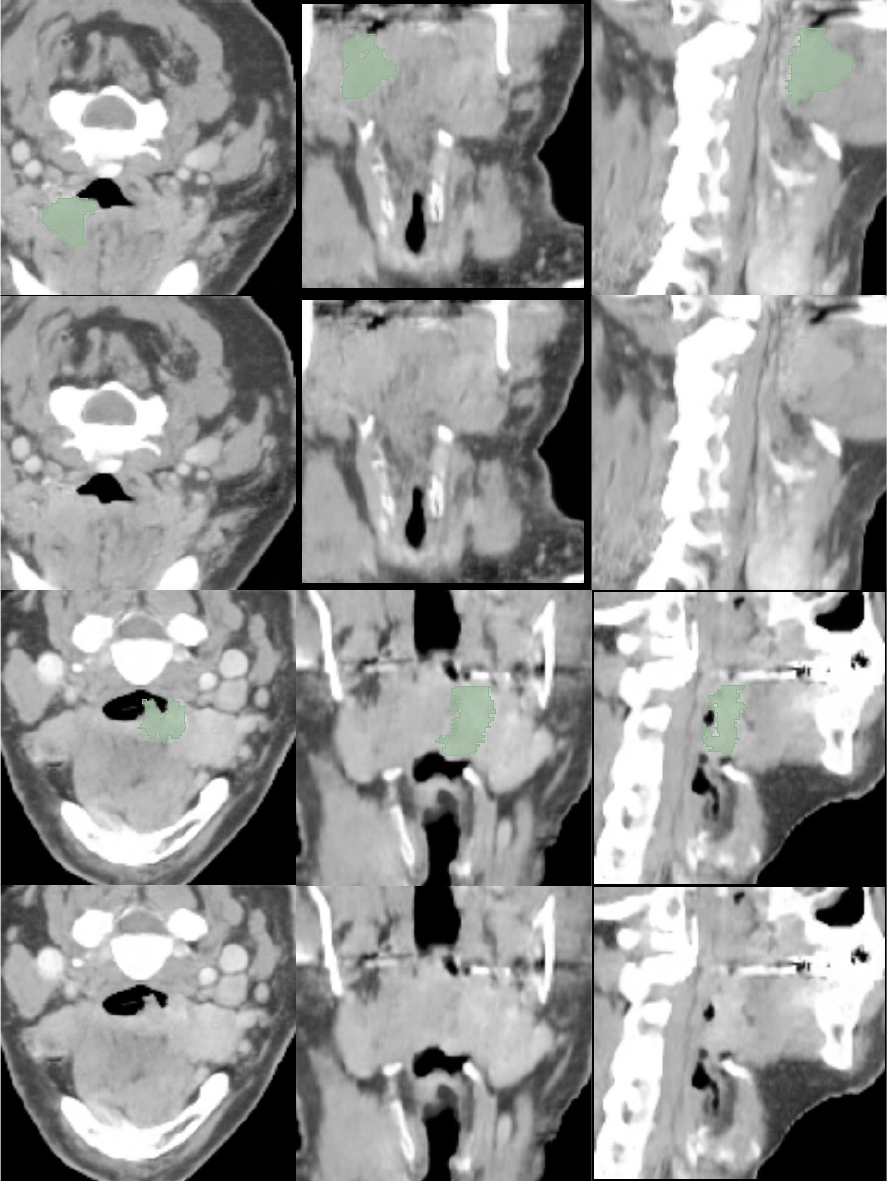}
\caption{Slices in the three axes of the synthetic CT scans generated using $DDPM_{all\_cat}^{200}$ and the sampling method DPM++ 2M SDE. The tumor is segmented in green.}
\label{fig:app:fake_tumour_cropped}
\end{figure}

\begin{figure}[h]
\centering
\includegraphics[width=0.8\linewidth]{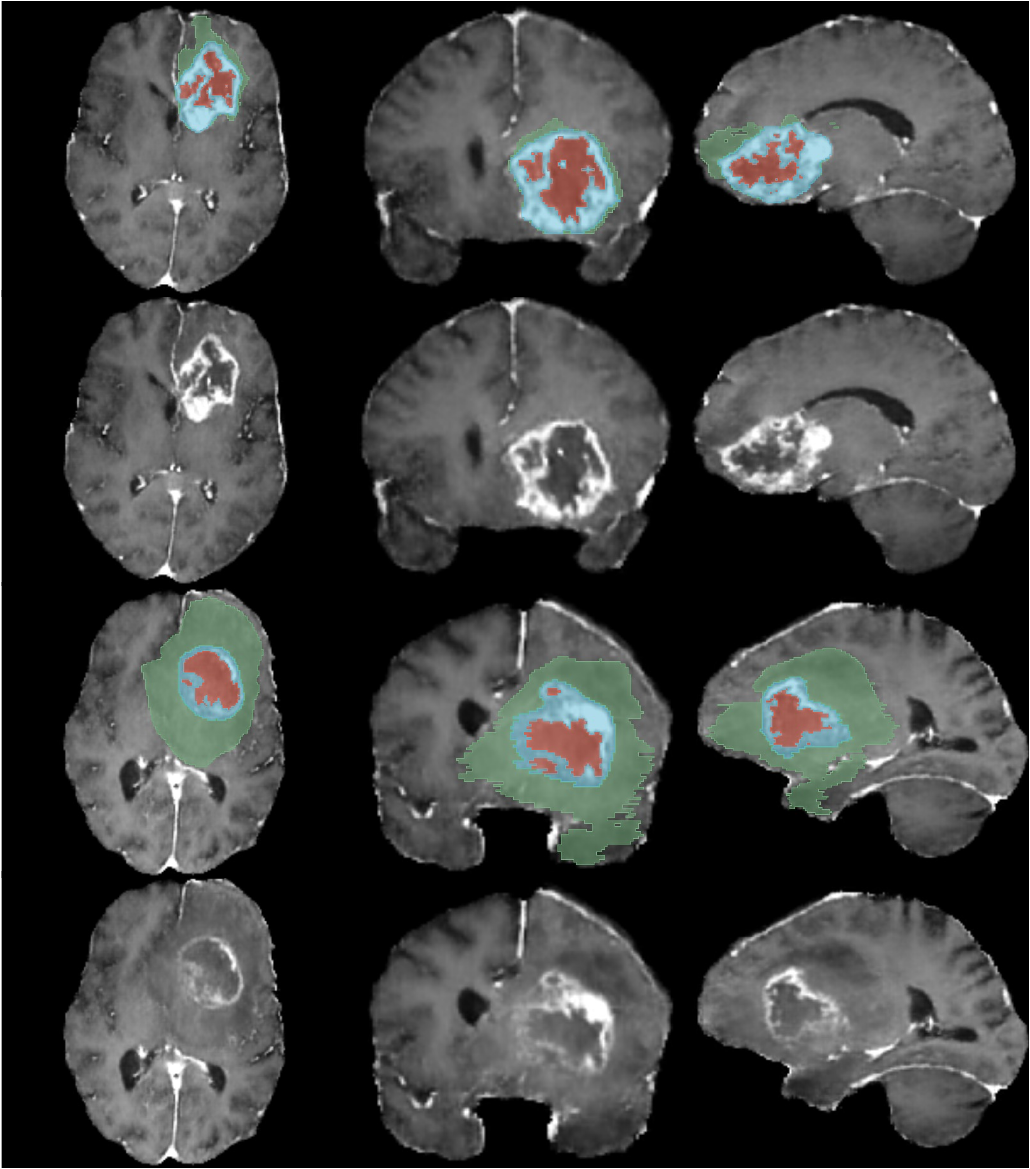}
\caption{Slices in the three axes of synthetic cases generated by the model $WDM_{seg\_conv}^{MRI}$ with the sampling method ‘DPM++ 2M SDE’. The first two rows were generated with the label BraTS-GLI-00000-000, and the last two with the label BraTS-GLI-00002-000. T1c with the necrotic tumor core (red), peritumoral edematous/invaded tissue (green)  and enhancing tumor (blue).}
\label{fig:app:fake_MRI}
\end{figure}

\begin{figure}[h]
\centering
\includegraphics[width=0.8\linewidth]{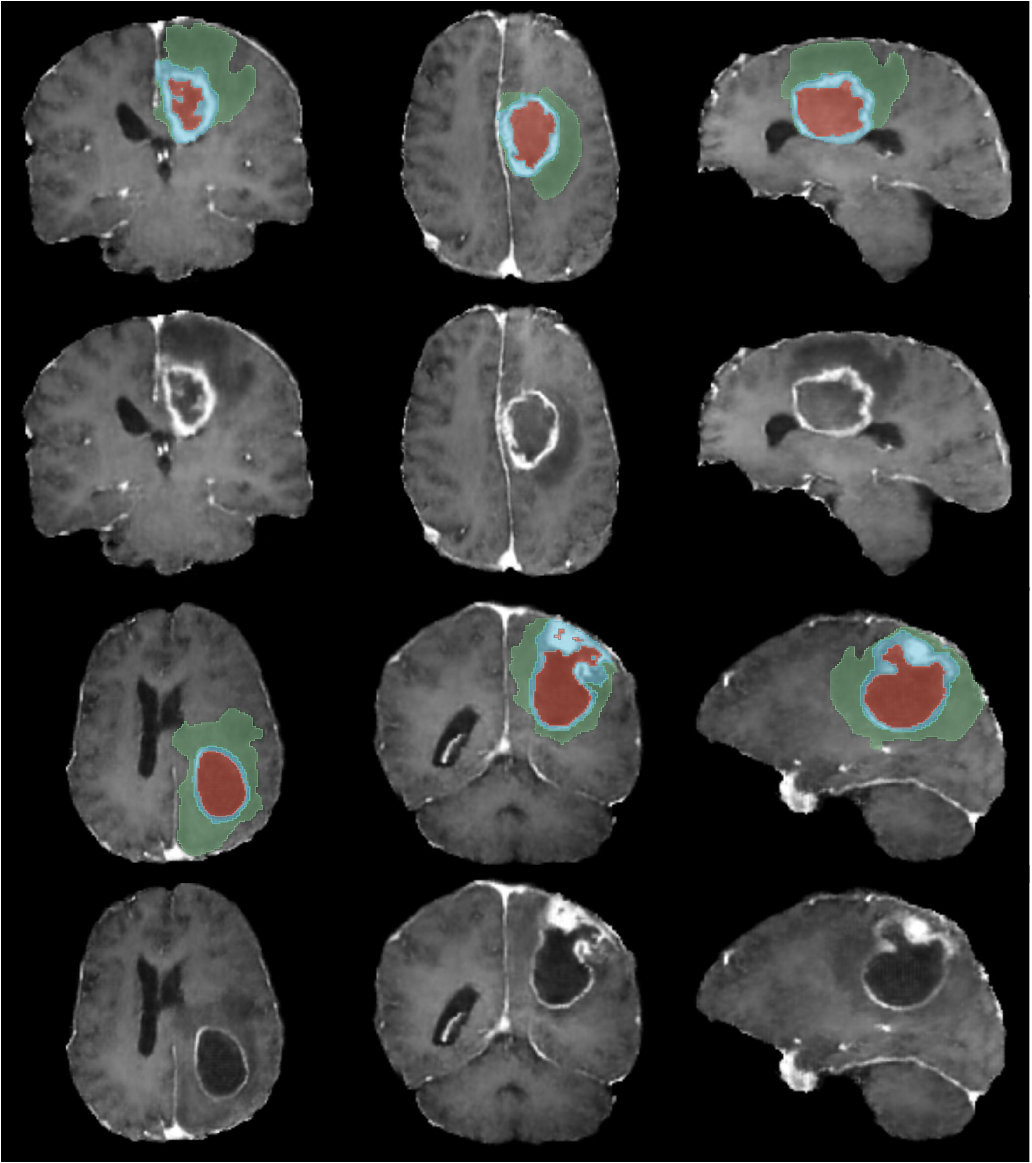}
\caption{Slices in the three axes of synthetic cases generated by the model $WDM_{seg\_conv}^{MRI}$ with the sampling method ‘DPM++ 2M SDE’. The first two rows were generated with the label BraTS-GLI-00003-000, and the last two with the label BraTS-GLI-00005-000. T1c with the necrotic tumor core (red), peritumoral edematous/invaded tissue (green)  and enhancing tumor (blue).}
\label{fig:app:fake_MRI_2}
\end{figure}

\begin{figure}[h]
\centering
\includegraphics[width=0.8\linewidth]{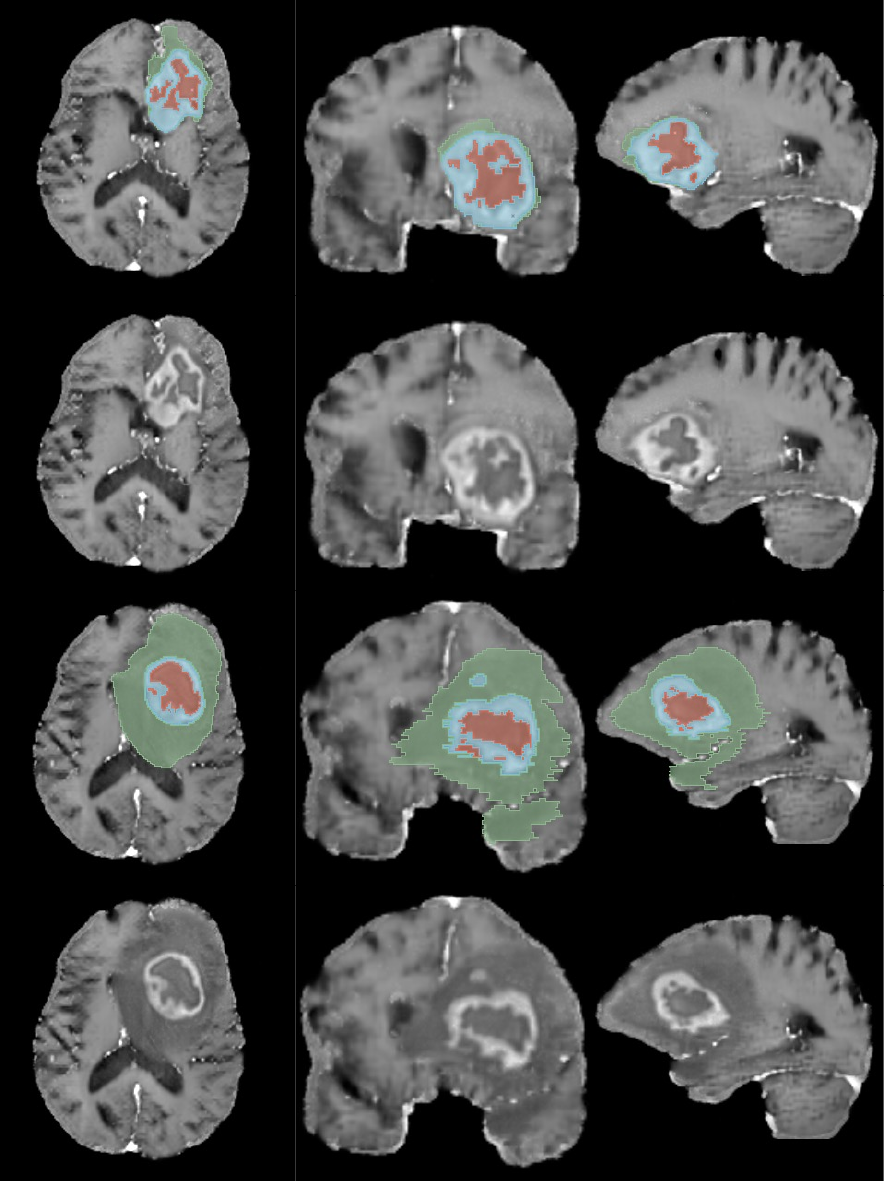}
\caption{Slices in the three axes of synthetic cases generated by the model $cGAN_{seg}^{MRI}$. The first two rows were generated with the label BraTS-GLI-00000-000, and the last two with the label BraTS-GLI-00002-000. T1c with the necrotic tumor core (red), peritumoral edematous/invaded tissue (green)  and enhancing tumor (blue).}
\label{fig:app:fake_MRI_GANs}
\end{figure}

In \ref{app:Synthetic_CT_scans_tumour} (Figures \ref{fig:app:fake_tumour_GAN}, \ref{fig:app:fake_tumour_convcat}, \ref{fig:app:fake_tumour_inpainted} and \ref{fig:app:fake_tumour_cropped}) and \ref{app:Synthetic_MRI_scans} (Figures \ref{fig:app:fake_MRI}, \ref{fig:app:fake_MRI_2} and \ref{fig:app:fake_MRI_GANs}) are presented CT and MRI synthetic cases with the respective tumor segmentations.

\begin{figure}[h]
\centering
\includegraphics[width=0.68\linewidth]{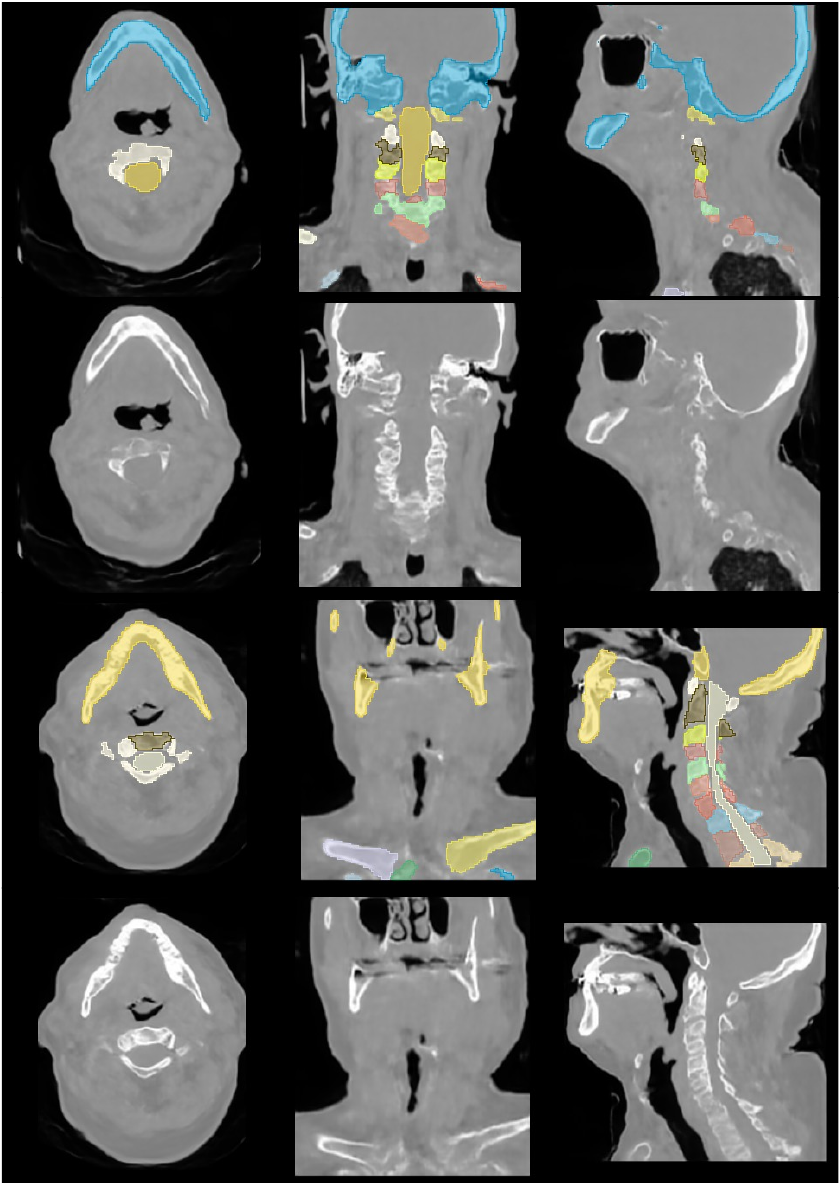}
\caption{Slices in the three axes of synthetic cases generated by the model $WDM_{ROI\_{d}}^{1000}$ with the sampling method ‘DPM++ 2M’ and the respective predictions of TotalSegmentator.}
\label{fig:app:fake_totalseg}
\end{figure}

\begin{figure}[h]
\centering
\includegraphics[width=0.7\linewidth]{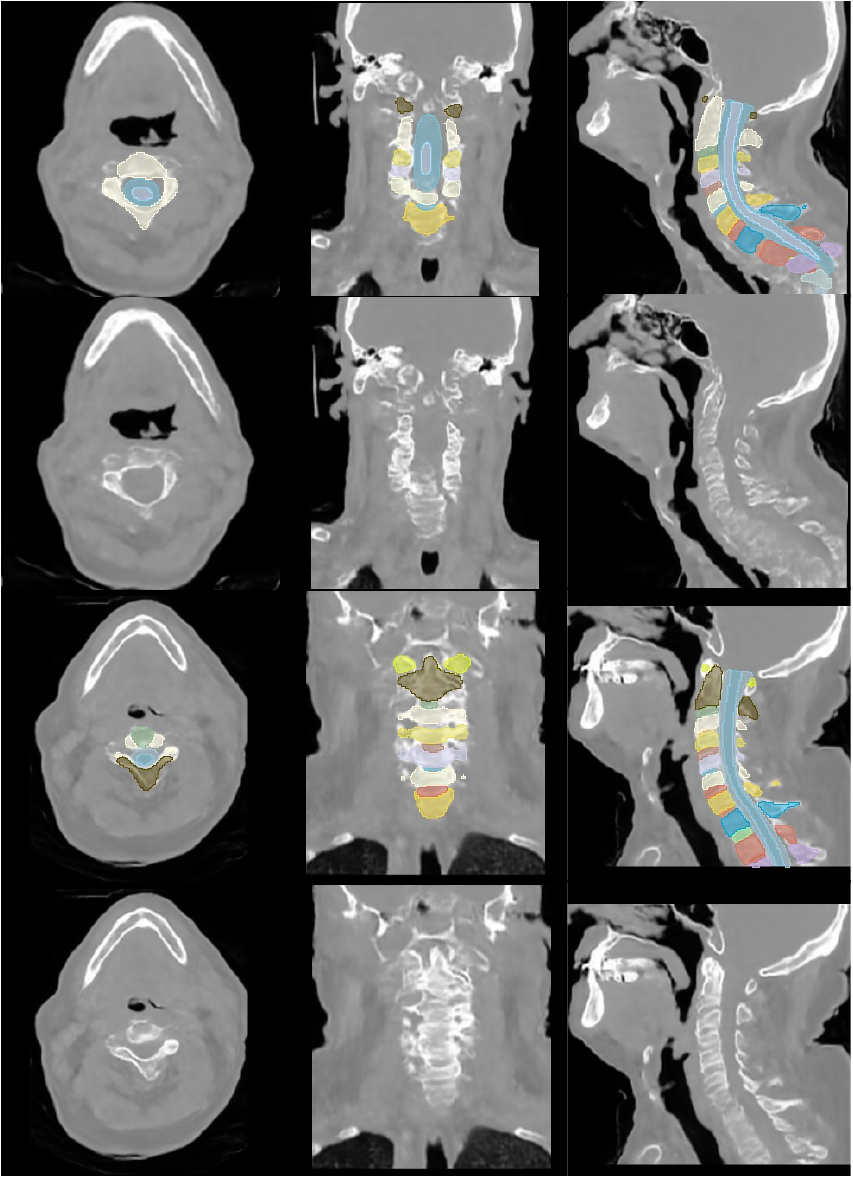}
\caption{Slices in the three axes of synthetic cases generated by the model $WDM_{ROI\_{d}}^{1000}$ with the sampling method ‘DPM++ 2M’ and the respective predictions of TotalSpineSeg.}
\label{fig:app:fake_spine}
\end{figure}

\begin{figure}[h]
\centering
\includegraphics[width=0.7\linewidth]{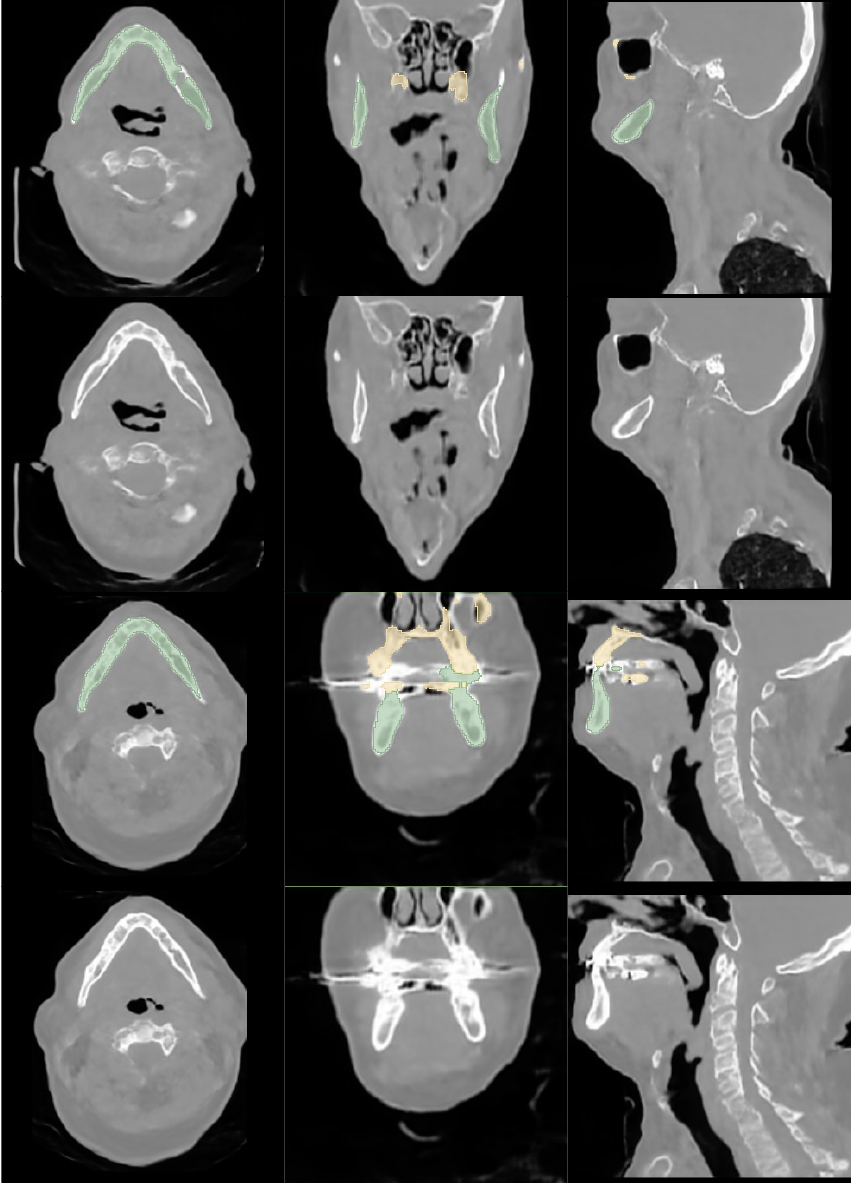}
\caption{Slices in the three axes of synthetic cases generated by the model $WDM_{ROI\_{d}}^{1000}$ with the sampling method ‘DPM++ 2M’ and the respective prediction of AMASSS-CBCT.}
\label{fig:app:fake_amasss}
\end{figure}

Figures \ref{fig:app:fake_totalseg}, \ref{fig:app:fake_spine} and \ref{fig:app:fake_amasss} in \ref{app:Synthetic_CT_scans_bone} present synthetic CT cases with the respective predictions of TotalSegmentator, TotalSpineSeg and AMASSS-CBCT.

\begin{table}[h]
\centering
\small
\setlength{\tabcolsep}{3pt} 
\renewcommand{\arraystretch}{1.5} 
\caption{Summary of experiments, including the models used, their respective tasks, and the modalities/dataset considered. The table outlines the relationship between each model, task, and structure to be segmented for MRI and CT modalities. The structures are the Whole Tumor (WT), Tumor Core (TC), Enhancing Tumor (ET), Gross Tumor Volume (GTV) and Bones.}
\begin{tabular}{@{}llll@{}}
\toprule
\textbf{Modality} & \textbf{Model} & \textbf{Task} & \textbf{Structures} \\
\midrule
\multirow{4}{*}{MRI ($MRI_{real}^{{1000}}$)}
  &  $cGAN_{seg}^{MRI}$ & Tumor segmentation & WT, TC, ET \\
  &  $WDM_{seg\_conv}^{MRI}$ & Tumor segmentation & WT, TC, ET \\
  &  $WDM_{seg\_d}^{MRI}$ & Tumor segmentation & WT, TC, ET \\
   & $WDM_{seg\_w}^{MRI}$ & Tumor segmentation & WT, TC, ET \\
\midrule
\multirow{6}{*}{CT ($CT_{real}^{1063}$)}
  & $cGAN_{all}^{200}$ & Tumor segmentation & GTV \\
  & $WDM_{all\_conv}^{200}$ & Tumor segmentation & GTV \\
  & $WDM_{all\_d}^{200}$ & Tumor segmentation & GTV \\
  & $WDM_{all\_w}^{200}$ & Tumor segmentation & GTV \\
  & $WDM_{ROI\_d}^{200}$ & Bone and Tumor segmentation & Bones, GTV \\
  & $WDM_{ROI\_d}^{1000}$ & Bone and Tumor segmentation & Bones, GTV \\
\midrule
\multirow{2}{*}{CT ($CT_{real}^{778}$)}
  & $DDPM_{all\_cat}^{200}$ & Tumor inpainting and segmentation & GTV \\
  & $DDPM_{all\_cat}^{1000}$ & Tumor inpainting and segmentation & GTV \\
\bottomrule
\end{tabular}%
\label{tab:exp_summary}
\end{table}

\begin{table}[h]
\small
\caption{Anatomical structures segmented by different bone segmentation tools}
\label{tab:DSC_bone_exp}
\renewcommand{\arraystretch}{1.0} 
\centering
\begin{tabular}{cccc}
\toprule
\textbf{Structures}        & \textbf{TotalSpineSeg} & \textbf{TotalSegmentator} & \textbf{AMASSS-CBCT} \\ 
\midrule

Skull                       &                        & \checkmark              &            \\ 
\cmidrule(lr){3-3}
Clavicula (L, R)           &                        & \checkmark              &                      \\
\cmidrule(lr){3-3}

Vertebrae (C1-T6)         & \checkmark             & \checkmark              &            \\ 
\cmidrule(lr){2-3}
Spinal discs (C2-T6) & \checkmark             &                          &                      \\ 
\cmidrule(lr){2-2}

Spinal cord                 & \checkmark             &                          &                      \\ 
\cmidrule(lr){2-2}
Spinal canal                & \checkmark             & \checkmark               &                      \\ 
\cmidrule(lr){2-3}

Ribs (L1-6, R1-6)           &                        & \checkmark              &                      \\ 
\cmidrule(lr){3-3}
Sternum                     &                        & \checkmark              &                      \\
\cmidrule(lr){3-3}

Costal cartilages           &                        & \checkmark              &                      \\ 
\cmidrule(lr){3-3}
Maxilla                     &                        &                          & \checkmark           \\ 
\cmidrule(lr){4-4}

Mandibula                   &                        &                          & \checkmark           \\ 
\cmidrule(lr){4-4}
\bottomrule
\end{tabular}
\end{table}

\begin{table}[h]
\centering
\small
\setlength{\tabcolsep}{3pt} 
\renewcommand{\arraystretch}{1.2} 
\caption{DSC results for tumor segmentation in H\&N CT, and MAE and MS-SSIM metrics comparing different sampling methods across four models and the real dataset. The best and second-best results for each metric are shown in bold and underlined.}
\begin{tabular}{@{}llccc@{}}
\toprule
\textbf{Model} & \textbf{Sampling Method} & \textbf{DSC} & \textbf{MAE} & \textbf{MS-SSIM}  \\
\midrule
Real & --- & 0.55345 & --- & 0.3581 ±0.1158 \\
\midrule
$cGAN_{all}^{200}$ & --- & 0.0083 & 0.2619 ±0.0874 & \textbf{0.3854} ±0.1074 \\
\midrule
\multirow{5}{*}{$WDM_{all\_conv}^{200}$} & DPM++ 2M & \textbf{0.0196} & 0.2153 ±0.0437 & \underline{0.3935} ±0.1081 \\
 & DPM++ 2M Karras & 0.0049 & 0.2162 ±0.0440 & 0.3940 ±0.1081 \\
 & DPM++ 2M SDE & \underline{0.0142} & 0.2142 ±0.0442 & 0.3970 ±0.1087 \\
 & DPM++ 2M SDE Karras & 0.0077 & 0.2162 ±0.0446 & 0.3974 ±0.1089 \\
 & Linear (1000) & 0.0018 & 0.2158 ±0.0447 & 0.3976 ±0.1087 \\
\midrule
\multirow{5}{*}{$WDM_{all\_d}^{200}$} & DPM++ 2M & 0.0025 & 0.2234 ±0.0465 & 0.3981 ±0.1096 \\
 & DPM++ 2M Karras & 0.0007 & 0.2260 ±0.0456 & 0.3983 ±0.1093 \\
 & DPM++ 2M SDE & 0.0026 & 0.2283 ±0.0461 & 0.3968 ±0.1094 \\
 & DPM++ 2M SDE Karras & 0.0004 & 0.2352 ±0.0440 & 0.3969 ±0.1094 \\
 & Linear (1000) & 0.0004 & 0.2304 ±0.0424 & 0.3866 ±0.1077 \\
\midrule
\multirow{5}{*}{$WDM_{all\_w}^{200}$} & DPM++ 2M & 0.0012 & 0.2135 ±0.0444 & 0.4069 ±0.1108 \\
 & DPM++ 2M Karras & 0.0009 & 0.2142 ±0.0445 & 0.4073 ±0.1109 \\
 & DPM++ 2M SDE & 0.0009 & \underline{0.2132} ±0.0443 & 0.4085 ±0.1112 \\
 & DPM++ 2M SDE Karras & 0.0020 & 0.2154 ±0.0442 & 0.4100 ±0.1114 \\
 & Linear (1000) & 0.0039 & \textbf{0.2127} ±0.0430 & 0.3976 ±0.1087 \\
\bottomrule
\end{tabular}%
\label{tab:ct_seg_and_metrics}
\end{table}

\begin{table}[h]
\centering
\small
\setlength{\tabcolsep}{3pt} 
\renewcommand{\arraystretch}{1.2} 
\caption{Aggregated DSC results for tumor segmentation on the MRI scans. The results are presented for each region and for the global segmentation. MAE and MS-SSIM metrics are also used for comparing different sampling methods across four models and the real dataset. The best and second-best results for each metric and region are shown in bold and underlined.}
\begin{tabular}{@{}llcccccc@{}}
\toprule
\multirow{2}{*}{\textbf{Model}} & \multirow{2}{*}{\shortstack{\textbf{Sampling}\\\textbf{Method}}} & \multicolumn{4}{c}{\textbf{DSC}} & \multirow{2}{*}{\textbf{MAE}} & \multirow{2}{*}{\textbf{MS-SSIM}} \\
\cmidrule(lr){3-6}
 &  & \textbf{WT} & \textbf{TC} & \textbf{ET} & \textbf{Mean} &  & \\

\midrule
\multirow{2}{*}{Real} & \multirow{2}{*}{---} & \multirow{2}{*}{\begin{tabular}[c]{@{}c@{}} 0.8667 \\ ±0.1201 \end{tabular}} & \multirow{2}{*}{\begin{tabular}[c]{@{}c@{}} 0.9171 \\ ±0.1359 \end{tabular}} & \multirow{2}{*}{\begin{tabular}[c]{@{}c@{}} 0.8853 \\ ±0.1534 \end{tabular}} & \multirow{2}{*}{\begin{tabular}[c]{@{}c@{}} 0.8897 \\ ±0.0208 \end{tabular}} & \multirow{2}{*}{---} & \multirow{2}{*}{\begin{tabular}[c]{@{}c@{}} 0.7201 \\ ±0.0594 \end{tabular}} \\
 &  &  &  &  &  &  &  \\
\midrule
\multirow{2}{*}{$cGAN_{seg}^{MRI}$} & \multirow{2}{*}{---} & \multirow{2}{*}{\begin{tabular}[c]{@{}c@{}} 0.6767 \\ ±0.2266 \end{tabular}} & \multirow{2}{*}{\begin{tabular}[c]{@{}c@{}} 0.7271 \\ ±0.2836 \end{tabular}} & \multirow{2}{*}{\begin{tabular}[c]{@{}c@{}} 0.6527 \\ ±0.2878 \end{tabular}} & \multirow{2}{*}{\begin{tabular}[c]{@{}c@{}} 0.6855 \\ ±0.031 \end{tabular}} & \multirow{2}{*}{\begin{tabular}[c]{@{}c@{}} 0.0805 \\ ±0.0308 \end{tabular}} & \multirow{2}{*}{\begin{tabular}[c]{@{}c@{}} 0.8551 \\ ±0.0655 \end{tabular}} \\
 &  &  &  &  &  &  &  \\
\midrule
\multirow{10}{*}{$WDM_{seg\_conv}^{MRI}$} & \multirow{2}{*}{DPM++ 2M} & \multirow{2}{*}{\begin{tabular}[c]{@{}c@{}} \underline{0.7636} \\ ±0.181 \end{tabular}} & \multirow{2}{*}{\begin{tabular}[c]{@{}c@{}} \underline{0.8923} \\ ±0.1633 \end{tabular}} & \multirow{2}{*}{\begin{tabular}[c]{@{}c@{}} 0.842 \\ ±0.18 \end{tabular}} & \multirow{2}{*}{\begin{tabular}[c]{@{}c@{}} \underline{0.8326} \\ ±0.053 \end{tabular}} & \multirow{2}{*}{\begin{tabular}[c]{@{}c@{}} 0.0254 \\ ±0.0058 \end{tabular}} & \multirow{2}{*}{\begin{tabular}[c]{@{}c@{}} 0.7071 \\ ±0.0665 \end{tabular}} \\
 &  &  &  &  &  &  &  \\
 & \multirow{2}{*}{\begin{tabular}[c]{@{}c@{}} DPM++ 2M \\ Karras \end{tabular}} & \multirow{2}{*}{\begin{tabular}[c]{@{}c@{}} 0.708 \\ ±0.1879 \end{tabular}} & \multirow{2}{*}{\begin{tabular}[c]{@{}c@{}} 0.8854 \\ ±0.1736 \end{tabular}} & \multirow{2}{*}{\begin{tabular}[c]{@{}c@{}} 0.841 \\ ±0.1779 \end{tabular}} & \multirow{2}{*}{\begin{tabular}[c]{@{}c@{}} 0.8115 \\ ±0.0754 \end{tabular}} & \multirow{2}{*}{\begin{tabular}[c]{@{}c@{}} 0.0237 \\ ±0.0053 \end{tabular}} & \multirow{2}{*}{\begin{tabular}[c]{@{}c@{}} 0.7071 \\ ±0.0677 \end{tabular}} \\
 &  &  &  &  &  &  &  \\
 & \multirow{2}{*}{\begin{tabular}[c]{@{}c@{}} DPM++ 2M \\ SDE \end{tabular}} & \multirow{2}{*}{\begin{tabular}[c]{@{}c@{}} \textbf{0.7654} \\ ±0.182 \end{tabular}} & \multirow{2}{*}{\begin{tabular}[c]{@{}c@{}} 0.8919 \\ ±0.1619 \end{tabular}} & \multirow{2}{*}{\begin{tabular}[c]{@{}c@{}} \underline{0.8452} \\ ±0.1733 \end{tabular}} & \multirow{2}{*}{\begin{tabular}[c]{@{}c@{}} \textbf{0.8342} \\ ±0.0522 \end{tabular}} & \multirow{2}{*}{\begin{tabular}[c]{@{}c@{}} 0.0252 \\ ±0.0058 \end{tabular}} & \multirow{2}{*}{\begin{tabular}[c]{@{}c@{}} 0.7064 \\ ±0.0675 \end{tabular}} \\
 &  &  &  &  &  &  &  \\
 & \multirow{2}{*}{\begin{tabular}[c]{@{}c@{}} DPM++ 2M \\ SDE Karras \end{tabular}} & \multirow{2}{*}{\begin{tabular}[c]{@{}c@{}} 0.6688 \\ ±0.2138 \end{tabular}} & \multirow{2}{*}{\begin{tabular}[c]{@{}c@{}} 0.8697 \\ ±0.2086 \end{tabular}} & \multirow{2}{*}{\begin{tabular}[c]{@{}c@{}} 0.8341 \\ ±0.1901 \end{tabular}} & \multirow{2}{*}{\begin{tabular}[c]{@{}c@{}} 0.7909 \\ ±0.0875 \end{tabular}} & \multirow{2}{*}{\begin{tabular}[c]{@{}c@{}} \textbf{0.0233} \\ ±0.0052 \end{tabular}} & \multirow{2}{*}{\begin{tabular}[c]{@{}c@{}} 0.7069 \\ ±0.0691 \end{tabular}} \\
 &  &  &  &  &  &  &  \\
 & \multirow{2}{*}{Linear (1000)} & \multirow{2}{*}{\begin{tabular}[c]{@{}c@{}} 0.6926 \\ ±0.1958 \end{tabular}} & \multirow{2}{*}{\begin{tabular}[c]{@{}c@{}} 0.8775 \\ ±0.1948 \end{tabular}} & \multirow{2}{*}{\begin{tabular}[c]{@{}c@{}} 0.8388 \\ ±0.1806 \end{tabular}} & \multirow{2}{*}{\begin{tabular}[c]{@{}c@{}} 0.803 \\ ±0.0796 \end{tabular}} & \multirow{2}{*}{\begin{tabular}[c]{@{}c@{}} \underline{0.0234} \\ ±0.0052 \end{tabular}} & \multirow{2}{*}{\begin{tabular}[c]{@{}c@{}} 0.7062 \\ ±0.0692 \end{tabular}} \\
 &  &  &  &  &  &  &  \\
\midrule
\multirow{10}{*}{$WDM_{seg\_d}^{MRI}$} & \multirow{2}{*}{DPM++ 2M} & \multirow{2}{*}{\begin{tabular}[c]{@{}c@{}} 0.722 \\ ±0.203 \end{tabular}} & \multirow{2}{*}{\begin{tabular}[c]{@{}c@{}} 0.8825 \\ ±0.1974 \end{tabular}} & \multirow{2}{*}{\begin{tabular}[c]{@{}c@{}} 0.8435 \\ ±0.1805 \end{tabular}} & \multirow{2}{*}{\begin{tabular}[c]{@{}c@{}} 0.816 \\ ±0.0683 \end{tabular}} & \multirow{2}{*}{\begin{tabular}[c]{@{}c@{}} 0.0292 \\ ±0.0063 \end{tabular}} & \multirow{2}{*}{\begin{tabular}[c]{@{}c@{}} 0.7079 \\ ±0.0646 \end{tabular}} \\
 &  &  &  &  &  &  &  \\
 & \multirow{2}{*}{\begin{tabular}[c]{@{}c@{}} DPM++ 2M \\ Karras \end{tabular}} & \multirow{2}{*}{\begin{tabular}[c]{@{}c@{}} 0.7058 \\ ±0.2045 \end{tabular}} & \multirow{2}{*}{\begin{tabular}[c]{@{}c@{}} 0.8731 \\ ±0.199 \end{tabular}} & \multirow{2}{*}{\begin{tabular}[c]{@{}c@{}} 0.8289 \\ ±0.1871 \end{tabular}} & \multirow{2}{*}{\begin{tabular}[c]{@{}c@{}} 0.8026 \\ ±0.0708 \end{tabular}} & \multirow{2}{*}{\begin{tabular}[c]{@{}c@{}} 0.0300 \\ ±0.0063 \end{tabular}} & \multirow{2}{*}{\begin{tabular}[c]{@{}c@{}} 0.7079 \\ ±0.0648 \end{tabular}} \\
 &  &  &  &  &  &  &  \\
 & \multirow{2}{*}{\begin{tabular}[c]{@{}c@{}} DPM++ 2M \\ SDE \end{tabular}} & \multirow{2}{*}{\begin{tabular}[c]{@{}c@{}} 0.7193 \\ ±0.2071 \end{tabular}} & \multirow{2}{*}{\begin{tabular}[c]{@{}c@{}} 0.8829 \\ ±0.1907 \end{tabular}} & \multirow{2}{*}{\begin{tabular}[c]{@{}c@{}} 0.8434 \\ ±0.1712 \end{tabular}} & \multirow{2}{*}{\begin{tabular}[c]{@{}c@{}} 0.8152 \\ ±0.0697 \end{tabular}} & \multirow{2}{*}{\begin{tabular}[c]{@{}c@{}} 0.0298 \\ ±0.0064 \end{tabular}} & \multirow{2}{*}{\begin{tabular}[c]{@{}c@{}} \underline{0.7081} \\ ±0.0647 \end{tabular}} \\
 &  &  &  &  &  &  &  \\
 & \multirow{2}{*}{\begin{tabular}[c]{@{}c@{}} DPM++ 2M \\ SDE Karras \end{tabular}} & \multirow{2}{*}{\begin{tabular}[c]{@{}c@{}} 0.6713 \\ ±0.2276 \end{tabular}} & \multirow{2}{*}{\begin{tabular}[c]{@{}c@{}} 0.84 \\ ±0.2468 \end{tabular}} & \multirow{2}{*}{\begin{tabular}[c]{@{}c@{}} 0.7955 \\ ±0.2281 \end{tabular}} & \multirow{2}{*}{\begin{tabular}[c]{@{}c@{}} 0.7689 \\ ±0.0714 \end{tabular}} & \multirow{2}{*}{\begin{tabular}[c]{@{}c@{}} 0.0320 \\ ±0.0064 \end{tabular}} & \multirow{2}{*}{\begin{tabular}[c]{@{}c@{}} \textbf{0.7092} \\ ±0.0639 \end{tabular}} \\
 &  &  &  &  &  &  &  \\
 & \multirow{2}{*}{Linear (1000)} & \multirow{2}{*}{\begin{tabular}[c]{@{}c@{}} 0.6966 \\ ±0.2065 \end{tabular}} & \multirow{2}{*}{\begin{tabular}[c]{@{}c@{}} 0.8638 \\ ±0.2128 \end{tabular}} & \multirow{2}{*}{\begin{tabular}[c]{@{}c@{}} 0.818 \\ ±0.1942 \end{tabular}} & \multirow{2}{*}{\begin{tabular}[c]{@{}c@{}} 0.7928 \\ ±0.0706 \end{tabular}} & \multirow{2}{*}{\begin{tabular}[c]{@{}c@{}} 0.0315 \\ ±0.0065 \end{tabular}} & \multirow{2}{*}{\begin{tabular}[c]{@{}c@{}} 0.7078 \\ ±0.0643 \end{tabular}} \\
 &  &  &  &  &  &  &  \\
\midrule
\multirow{10}{*}{$WDM_{seg\_w}^{MRI}$} & \multirow{2}{*}{DPM++ 2M} & \multirow{2}{*}{\begin{tabular}[c]{@{}c@{}} 0.7103 \\ ±0.2072 \end{tabular}} & \multirow{2}{*}{\begin{tabular}[c]{@{}c@{}} \textbf{0.8999} \\ ±0.1333 \end{tabular}} & \multirow{2}{*}{\begin{tabular}[c]{@{}c@{}} \textbf{0.8453} \\ ±0.1573 \end{tabular}} & \multirow{2}{*}{\begin{tabular}[c]{@{}c@{}} 0.8185 \\ ±0.0797 \end{tabular}} & \multirow{2}{*}{\begin{tabular}[c]{@{}c@{}} 0.0292 \\ ±0.0059 \end{tabular}} & \multirow{2}{*}{\begin{tabular}[c]{@{}c@{}} 0.7065 \\ ±0.0640 \end{tabular}} \\
 &  &  &  &  &  &  &  \\
 & \multirow{2}{*}{\begin{tabular}[c]{@{}c@{}} DPM++ 2M \\ Karras \end{tabular}} & \multirow{2}{*}{\begin{tabular}[c]{@{}c@{}} 0.6998 \\ ±0.1916 \end{tabular}} & \multirow{2}{*}{\begin{tabular}[c]{@{}c@{}} 0.8875 \\ ±0.1679 \end{tabular}} & \multirow{2}{*}{\begin{tabular}[c]{@{}c@{}} 0.832 \\ ±0.1814 \end{tabular}} & \multirow{2}{*}{\begin{tabular}[c]{@{}c@{}} 0.8064 \\ ±0.0788 \end{tabular}} & \multirow{2}{*}{\begin{tabular}[c]{@{}c@{}} 0.0309 \\ ±0.0059 \end{tabular}} & \multirow{2}{*}{\begin{tabular}[c]{@{}c@{}} 0.7057 \\ ±0.0642 \end{tabular}} \\
 &  &  &  &  &  &  &  \\
 & \multirow{2}{*}{\begin{tabular}[c]{@{}c@{}} DPM++ 2M \\ SDE \end{tabular}} & \multirow{2}{*}{\begin{tabular}[c]{@{}c@{}} 0.7174 \\ ±0.2033 \end{tabular}} & \multirow{2}{*}{\begin{tabular}[c]{@{}c@{}} 0.8882 \\ ±0.1664 \end{tabular}} & \multirow{2}{*}{\begin{tabular}[c]{@{}c@{}} 0.8373 \\ ±0.1707 \end{tabular}} & \multirow{2}{*}{\begin{tabular}[c]{@{}c@{}} 0.8143 \\ ±0.0716 \end{tabular}} & \multirow{2}{*}{\begin{tabular}[c]{@{}c@{}} 0.0299 \\ ±0.0059 \end{tabular}} & \multirow{2}{*}{\begin{tabular}[c]{@{}c@{}} 0.7067 \\ ±0.0637 \end{tabular}} \\
 &  &  &  &  &  &  &  \\
 & \multirow{2}{*}{\begin{tabular}[c]{@{}c@{}} DPM++ 2M \\ SDE Karras \end{tabular}} & \multirow{2}{*}{\begin{tabular}[c]{@{}c@{}} 0.6672 \\ ±0.1991 \end{tabular}} & \multirow{2}{*}{\begin{tabular}[c]{@{}c@{}} 0.8762 \\ ±0.1841 \end{tabular}} & \multirow{2}{*}{\begin{tabular}[c]{@{}c@{}} 0.8294 \\ ±0.1706 \end{tabular}} & \multirow{2}{*}{\begin{tabular}[c]{@{}c@{}} 0.7909 \\ ±0.0895 \end{tabular}} & \multirow{2}{*}{\begin{tabular}[c]{@{}c@{}} 0.0324 \\ ±0.0059 \end{tabular}} & \multirow{2}{*}{\begin{tabular}[c]{@{}c@{}} 0.7062 \\ ±0.0635 \end{tabular}} \\
 &  &  &  &  &  &  &  \\
 & \multirow{2}{*}{Linear (1000)} & \multirow{2}{*}{\begin{tabular}[c]{@{}c@{}} 0.6661 \\ ±0.2014 \end{tabular}} & \multirow{2}{*}{\begin{tabular}[c]{@{}c@{}} 0.8798 \\ ±0.1821 \end{tabular}} & \multirow{2}{*}{\begin{tabular}[c]{@{}c@{}} 0.8374 \\ ±0.1619 \end{tabular}} & \multirow{2}{*}{\begin{tabular}[c]{@{}c@{}} 0.7945 \\ ±0.0924 \end{tabular}} & \multirow{2}{*}{\begin{tabular}[c]{@{}c@{}} 0.0320 \\ ±0.0058 \end{tabular}} & \multirow{2}{*}{\begin{tabular}[c]{@{}c@{}} 0.7052 \\ ±0.0637 \end{tabular}} \\
 &  &  &  &  &  &  &  \\
\bottomrule
\end{tabular}%
\label{tab:mri_seg_results}
\end{table}

\begin{table}[h]
\centering
\small
\setlength{\tabcolsep}{3pt} 
\caption{DSC results for tumor segmentation in H\&N CT comparing different sampling methods across four models and inpainting combination. The first two datasets are full resolution cases with inpainted tumor. The last two datasets are random patches generated with the inpainted model used directly as input of the nnU-Net. The best and second-best results for each method are shown in bold and underlined}
\begin{tabular}{@{}llcc@{}}
\toprule
\multirow{2}{*}{\textbf{Model}} & \multirow{2}{*}{\shortstack{\textbf{Sampling}\\\textbf{Method}}} & \multicolumn{2}{c}{\textbf{DSC}} \\
\cmidrule(lr){3-4}
  &  & \textbf{Edge blur} & \textbf{Full blur} \\
\midrule
\multirow{5}{*}{$WDM_{ROI\_d}^{200}$ + $DDPM_{all\_{cat}}^{200}$}
  & DPM++ 2M            & 0.0189          & 0.0133          \\
  & DPM++ 2M Karras     & 0.0160          & 0.0081          \\
  & DPM++ 2M SDE        & 0.0202          & 0.0095          \\
  & DPM++ 2M SDE Karras & 0.0211          & 0.0020          \\
  & Linear (1000)       & 0.0181          & \textbf{0.0324} \\
\midrule
\multirow{5}{*}{$WDM_{ROI\_d}^{1000}$ + $DDPM_{all\_{cat}}^{1000}$}
  & DPM++ 2M            & 0.0182          & 0.0191          \\
  & DPM++ 2M Karras     & 0.0276          & 0.0149          \\
  & DPM++ 2M SDE        & 0.0205          & \underline{0.0241} \\
  & DPM++ 2M SDE Karras & \textbf{0.0355} & 0.0174          \\
  & Linear (1000)       & \underline{0.0331} & 0.0037       \\
\midrule
\multirow{5}{*}{$DDPM_{all\_cat}^{200}$}
  & DPM++ 2M            & \multicolumn{2}{c}{\underline{0.0559}} \\
  & DPM++ 2M Karras     & \multicolumn{2}{c}{0.0057} \\
  & DPM++ 2M SDE        & \multicolumn{2}{c}{\textbf{0.0640}} \\
  & DPM++ 2M SDE Karras & \multicolumn{2}{c}{0.0025} \\
  & Linear (1000)       & \multicolumn{2}{c}{0.0027} \\
\midrule
\multirow{5}{*}{$DDPM_{all\_cat}^{1000}$}
  & DPM++ 2M            & \multicolumn{2}{c}{0.0072} \\
  & DPM++ 2M Karras     & \multicolumn{2}{c}{0.0117} \\
  & DPM++ 2M SDE        & \multicolumn{2}{c}{0.0241} \\
  & DPM++ 2M SDE Karras & \multicolumn{2}{c}{0.0100} \\
  & Linear (1000)       & \multicolumn{2}{c}{0.0094} \\
\bottomrule
\end{tabular}%
\label{tab:inpainted_dsc}
\end{table}

\begin{table}[h]
\centering
\small
\setlength{\tabcolsep}{3pt} 
\caption{MAE and MS-SSIM metrics for H\&N CT scans used for bone segmentation, comparing different sampling methods across two models. The best and second-best results for each metric are shown in bold and underlined}
\begin{tabular}{@{}llcc@{}}
\toprule
\multirow{2}{*}{\textbf{Model}} & \multirow{2}{*}{\textbf{Sampling Method}} & \multicolumn{2}{c}{\textbf{Metrics}} \\
\cmidrule(lr){3-4}
  &  & \textbf{MAE} & \textbf{MS-SSIM} \\
\midrule
\multirow{5}{*}{$WDM_{ROI\_d}^{200}$}
  & DPM++ 2M            & 0.3684 ± 0.0651 & \underline{0.3560} ± 0.1140 \\
  & DPM++ 2M Karras     & 0.3710 ± 0.0669 & 0.3552 ± 0.1134 \\
  & DPM++ 2M SDE        & 0.3486 ± 0.0648 & 0.3762 ± 0.1193 \\
  & DPM++ 2M SDE Karras & 0.3564 ± 0.0645 & 0.3757 ± 0.1183 \\
  & Linear (1000)       & 0.3512 ± 0.0639 & 0.3663 ± 0.1180 \\
\midrule
\multirow{5}{*}{$WDM_{ROI\_d}^{1000}$}
  & DPM++ 2M            & 0.2732 ± 0.0599 & \textbf{0.3586} ± 0.1121 \\
  & DPM++ 2M Karras     & 0.2737 ± 0.0572 & 0.3604 ± 0.1121 \\
  & DPM++ 2M SDE        & \textbf{0.2631} ± 0.0548 & 0.3723 ± 0.1139 \\
  & DPM++ 2M SDE Karras & \underline{0.2664} ± 0.0579 & 0.3750 ± 0.1153 \\
  & Linear (1000)       & 0.4872 ± 0.0438 & 0.4777 ± 0.1203 \\
\bottomrule
\end{tabular}%
\label{tab:ct_metrics}
\end{table}

\begin{table}[h]
\centering
\small
\setlength{\tabcolsep}{3pt} 
\caption{DSC results for bones segmentation on H\&N CT scans. Semantic and instance segmentations are tested for both models ($WDM_{ROI\_d}^{200}$ and $WDM_{ROI\_d}^{1000}$) comparing different sampling methods. The results for the real dataset are presented in the end}
\begin{tabular}{@{}llcccll@{}}
\toprule
\multirow{2}{*}{\shortstack{\textbf{Sampling}\\\textbf{Method}}} & \multirow{2}{*}{\textbf{Pre-segmentor}} & \multirow{2}{*}{\textbf{Structure}} & \multicolumn{2}{c}{\textbf{semantic DSC}} & \multicolumn{2}{c}{\textbf{instance DSC}} \\
\cmidrule(lr){4-5} \cmidrule(lr){6-7}
  &  &  & $WDM_{ROI\_d}^{200}$ & $WDM_{ROI\_d}^{1000}$ & $WDM_{ROI\_d}^{200}$ & $WDM_{ROI\_d}^{1000}$ \\
\midrule
\multirow{13}{*}{\shortstack{\textbf{DPM++}\\\textbf{2M}}} 
  & \multirow{4}{*}{TotalSpineSeg} 
    & Vertebrae & 0.9258 & \underline{0.9323} & \underline{0.9209} & 0.9182 \\
  &  & Spinal discs & 0.8884 & \underline{0.8984} & 0.8870 & 0.8914 \\
  &  & Spinal canal & 0.9213 & \textbf{0.9292} & 0.9213 & \textbf{0.9292} \\
  &  & Spinal cord & 0.9365 & \textbf{0.9413} & 0.9365 & \textbf{0.9413} \\
\cmidrule(lr){2-7}
  & \multirow{7}{*}{TotalSegmentator}
    & Skull & 0.7879 & \textbf{0.9524} & 0.7879 & \textbf{0.9524} \\
  &  & Clavicula & \textbf{0.7721} & \underline{0.9285} & 0.0406 & \underline{0.1439} \\
  &  & Vertebrae & 0.8273 & \underline{0.9372} & 0.8176 & \underline{0.9280} \\
  &  & Ribs & 0.8449 & \underline{0.8772} & 0.0294 & \textbf{0.2646} \\
  &  & Sternum & 0.9265 & \textbf{0.9469} & 0.9265 & \textbf{0.9469} \\
  &  & Costal cartilages & 0.5814 & 0.7883 & 0.5814 & 0.7883 \\
  &  & Spinal canal & 0.9045 & \textbf{0.9416} & 0.9045 & \textbf{0.9416} \\
\cmidrule(lr){2-7}
  & \multirow{2}{*}{AMASSS-CBCT}
    & Maxilla & --- & \textbf{0.8748} & --- & \textbf{0.8748} \\
  &  & Mandibula & --- & \textbf{0.9658} & --- & \textbf{0.9658} \\
\midrule
\multirow{13}{*}{\shortstack{\textbf{DPM++}\\\textbf{2M}\\\textbf{Karras}}} 
  & \multirow{4}{*}{TotalSpineSeg}
    & Vertebrae & \textbf{0.9274} & 0.9318 & \textbf{0.9221} & \textbf{0.9288} \\
  &  & Spinal discs & 0.8904 & 0.8921 & \underline{0.8888} & \underline{0.8921} \\
  &  & Spinal canal & \underline{0.9235} & 0.9276 & \underline{0.9235} & 0.9276 \\
  &  & Spinal cord & \underline{0.9378} & 0.9396 & \underline{0.9378} & 0.9396 \\
\cmidrule(lr){2-7}
  & \multirow{7}{*}{TotalSegmentator}
    & Skull & 0.7848 & \underline{0.9522} & 0.7848 & \underline{0.9522} \\
  &  & Clavicula & \underline{0.7587} & \textbf{0.9401} & 0.0469 & 0.1425 \\
  &  & Vertebrae & 0.8302 & \textbf{0.9373} & 0.8203 & \textbf{0.9285} \\
  &  & Ribs & 0.8428 & \textbf{0.8897} & 0.0352 & \underline{0.2349} \\
  &  & Sternum & 0.9284 & \underline{0.9453} & 0.9284 & \underline{0.9453} \\
  &  & Costal cartilages & \underline{0.5860} & \textbf{0.7971} & \underline{0.5860} & \textbf{0.7971} \\
  &  & Spinal canal & \underline{0.9048} & \underline{0.9411} & \underline{0.9048} & \underline{0.9411} \\
\cmidrule(lr){2-7}
  & \multirow{2}{*}{AMASSS-CBCT}
    & Maxilla & --- & \underline{0.8736} & --- & \underline{0.8736} \\
  &  & Mandibula & --- & \textbf{0.9658} & --- & \textbf{0.9658} \\
\midrule
\multirow{13}{*}{\shortstack{\textbf{DPM++}\\\textbf{2M}\\\textbf{SDE}}} 
  & \multirow{4}{*}{TotalSpineSeg}
    & Vertebrae & \underline{0.9269} & \textbf{0.9337} & 0.9199 & \underline{0.9261} \\
  &  & Spinal discs & \textbf{0.8925} & \textbf{0.8995} & \textbf{0.8891} & \textbf{0.8961} \\
  &  & Spinal canal & \textbf{0.9240} & \underline{0.9290} & \textbf{0.9240} & \underline{0.9289} \\
  &  & Spinal cord & \textbf{0.9386} & \underline{0.9405} & \textbf{0.9386} & \underline{0.9405} \\
\cmidrule(lr){2-7}
  & \multirow{7}{*}{TotalSegmentator}
    & Skull & \textbf{0.8000} & 0.9517 & \textbf{0.8000} & 0.9517 \\
  &  & Clavicula & 0.7435 & 0.7135 & 0.0398 & 0.0714 \\
  &  & Vertebrae & 0.8288 & 0.9361 & 0.8191 & 0.9230 \\
  &  & Ribs & 0.8413 & 0.7066 & 0.0286 & 0.1568 \\
  &  & Sternum & 0.9234 & 0.9435 & 0.9234 & 0.9435 \\
  &  & Costal cartilages & \textbf{0.6026} & 0.7586 & \textbf{0.6026} & 0.7586 \\
  &  & Spinal canal & 0.9030 & 0.9401 & 0.9030 & 0.9401 \\
\cmidrule(lr){2-7}
  & \multirow{2}{*}{AMASSS-CBCT}
    & Maxilla & --- & 0.8716 & --- & 0.8716 \\
  &  & Mandibula & --- & \underline{0.9657} & --- & \underline{0.9657} \\
\bottomrule
\end{tabular}
\label{tab:bone_seg_results}
\end{table}

\begin{table}[h]
\centering
\small
\setlength{\tabcolsep}{3pt} 
\caption{(Continuation) DSC results for bones segmentation on H\&N CT scans. Semantic and instance segmentations are tested for both models ($WDM_{ROI\_d}^{200}$ and $WDM_{ROI\_d}^{1000}$) comparing different sampling methods. The results for the real dataset are presented in the end}
\begin{tabular}{@{}llcccll@{}}
\toprule
\multirow{2}{*}{\shortstack{\textbf{Sampling}\\\textbf{Method}}} & \multirow{2}{*}{\textbf{Pre-segmentor}} & \multirow{2}{*}{\textbf{Structure}} & \multicolumn{2}{c}{\textbf{semantic DSC}} & \multicolumn{2}{c}{\textbf{instance DSC}} \\
\cmidrule(lr){4-5} \cmidrule(lr){6-7}
  &  &  & $WDM_{ROI\_d}^{200}$ & $WDM_{ROI\_d}^{1000}$ & $WDM_{ROI\_d}^{200}$ & $WDM_{ROI\_d}^{1000}$ \\
\midrule
\multirow{13}{*}{\shortstack{\textbf{DPM++}\\\textbf{2M}\\\textbf{SDE}\\\textbf{Karras}}} 
  & \multirow{4}{*}{TotalSpineSeg} 
    & Vertebrae (C1-T6) & 0.9248 & 0.9317 & 0.9192 & 0.9214 \\
  &  & Spinal discs (C2-T6) & 0.8865 & 0.8954 & 0.8840 & 0.8907 \\
  &  & Spinal canal & 0.9190 & 0.9268 & 0.9190 & 0.9268 \\
  &  & Spinal cord & 0.9346 & 0.9390 & 0.9346 & 0.9390 \\
  & \multirow{7}{*}{TotalSegmentator} 
    & Skull & \underline{0.7975} & 0.9519 & \underline{0.7975} & 0.9519 \\
  &  & Clavicula (L,R) & 0.7376 & 0.9123 & \underline{0.0551} & \textbf{0.1481} \\
  &  & Vertebrae (C1-T6) & \underline{0.8388} & 0.9365 & \underline{0.8288} & 0.9271 \\
  &  & Ribs (L1-6, R1-6) & \textbf{0.8469} & 0.8371 & \underline{0.0561} & 0.1505 \\
  &  & Sternum & \textbf{0.9296} & 0.9443 & \textbf{0.9296} & 0.9443 \\
  &  & Costal cartilages & 0.5779 & \underline{0.7946} & 0.5779 & \underline{0.7946} \\
  &  & Spinal canal & 0.9039 & 0.9406 & 0.9039 & 0.9406 \\
  & \multirow{2}{*}{AMASSS-CBCT} 
    & Maxilla & --- & 0.8719 & --- & 0.8719\\
  &  & Mandibula & --- & 0.9646 & --- & 0.9646 \\
\midrule
\multirow{13}{*}{\shortstack{\textbf{Linear}\\\textbf{(1000)}}} 
  & \multirow{4}{*}{TotalSpineSeg} 
    & Vertebrae (C1-T6) & 0.9243 & 0.9019 & 0.9169 & 0.8872 \\
  &  & Spinal discs (C2-T6) & 0.8710 & 0.8415 & 0.8689 & 0.8335 \\
  &  & Spinal canal & 0.9176 & 0.9066 & 0.9176 & 0.9066 \\
  &  & Spinal cord & 0.9320 & 0.9232 & 0.9320 & 0.9232 \\
\cmidrule(lr){2-7}
  & \multirow{7}{*}{TotalSegmentator} 
    & Skull & 0.7969 & 0.4932 & 0.7969 & 0.4932 \\
  &  & Clavicula (L,R) & 0.7402 & 0.1409 & \textbf{0.0674} & 0.0397 \\
  &  & Vertebrae (C1-T6) & \textbf{0.8401} & 0.5889 & \textbf{0.8300} & 0.5557 \\
  &  & Ribs (L1-6, R1-6) & \underline{0.8466} & 0.2285 & \textbf{0.0614} & 0.0705 \\
  &  & Sternum & \underline{0.9287} & 0.2784 & \underline{0.9287} & 0.2784 \\
  &  & Costal cartilages & 0.5763 & 0.0002 & 0.5763 & 0.0002 \\
  &  & Spinal canal & \textbf{0.9075} & 0.8540 & \textbf{0.9075} & 0.8540 \\
  \cmidrule(lr){2-7}
  & \multirow{2}{*}{AMASSS-CBCT} 
    & Maxilla & --- & 0.6755 & --- & 0.6755 \\
  &  & Mandibula & --- & 0.7552 & --- & 0.7552  \\
\midrule
\multirow{13}{*}{\textbf{Real}} 
  & \multirow{4}{*}{TotalSpineSeg} 
    & Vertebrae (C1-T6) & \multicolumn{2}{c}{0.9587} & \multicolumn{2}{c}{0.9575} \\
  &  & Spinal discs (C2-T6) & \multicolumn{2}{c}{0.9435} & \multicolumn{2}{c}{0.9435} \\
  &  & Spinal canal & \multicolumn{2}{c}{0.9508} & \multicolumn{2}{c}{0.9508} \\
  &  & Spinal cord & \multicolumn{2}{c}{0.9601} & \multicolumn{2}{c}{0.9601} \\
\cmidrule(lr){2-7}
  & \multirow{7}{*}{TotalSegmentator} 
    & Skull & \multicolumn{2}{c}{0.9536} & \multicolumn{2}{c}{0.9536} \\
  &  & Clavicula & \multicolumn{2}{c}{0.9488} & \multicolumn{2}{c}{0.9409} \\
  &  & Vertebrae & \multicolumn{2}{c}{0.9383} & \multicolumn{2}{c}{0.9311} \\
  &  & Ribs & \multicolumn{2}{c}{0.8999} & \multicolumn{2}{c}{0.8870} \\
  &  & Sternum & \multicolumn{2}{c}{0.9501} & \multicolumn{2}{c}{0.9501} \\
  &  & Costal Cartilages & \multicolumn{2}{c}{0.8455} & \multicolumn{2}{c}{0.8455} \\
  &  & Spinal Canal & \multicolumn{2}{c}{0.9453} & \multicolumn{2}{c}{0.9453} \\
\cmidrule(lr){2-7}
  & \multirow{2}{*}{AMASSS-CBCT} 
    & Maxilla & \multicolumn{2}{c}{0.8916} & \multicolumn{2}{c}{0.8915} \\
  &  & Mandibula & \multicolumn{2}{c}{0.9682} & \multicolumn{2}{c}{0.9682} \\
\bottomrule
\end{tabular}%
\label{tab:bone_seg_results_2}
\end{table}

\clearpage
\section{Real CT scans with tumor}
\label{app:Real_CT_scans}
\begin{figure}[h]
\centering
\includegraphics[width=0.8\linewidth]{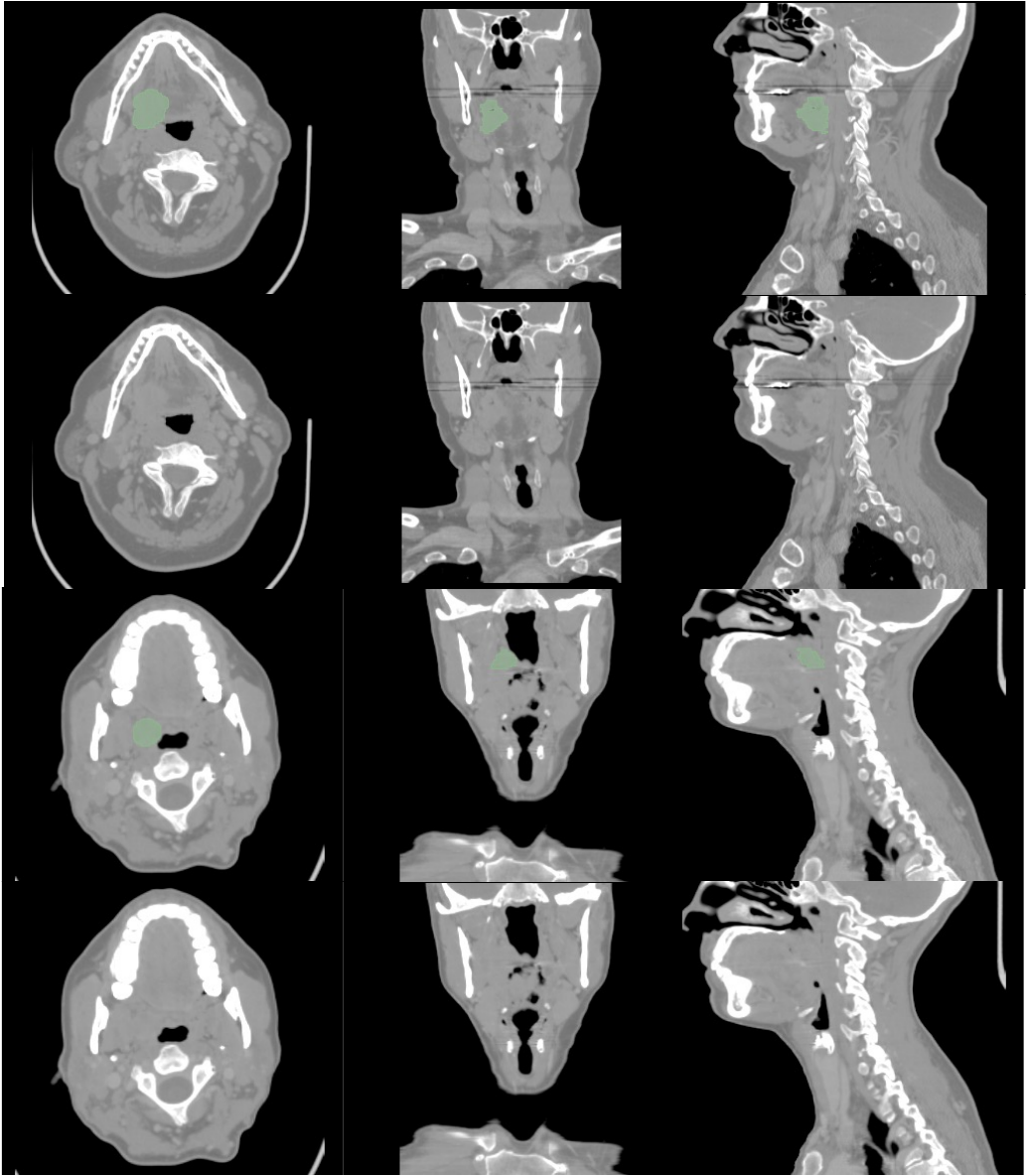}
\caption{Slices in the three axes of the real cases from $CT_{real}^{1258}$. The tumor is segmented in green.}
\label{fig:app:real_tumour}
\end{figure} 

\begin{figure}[h]
\centering
\includegraphics[width=0.8\linewidth]{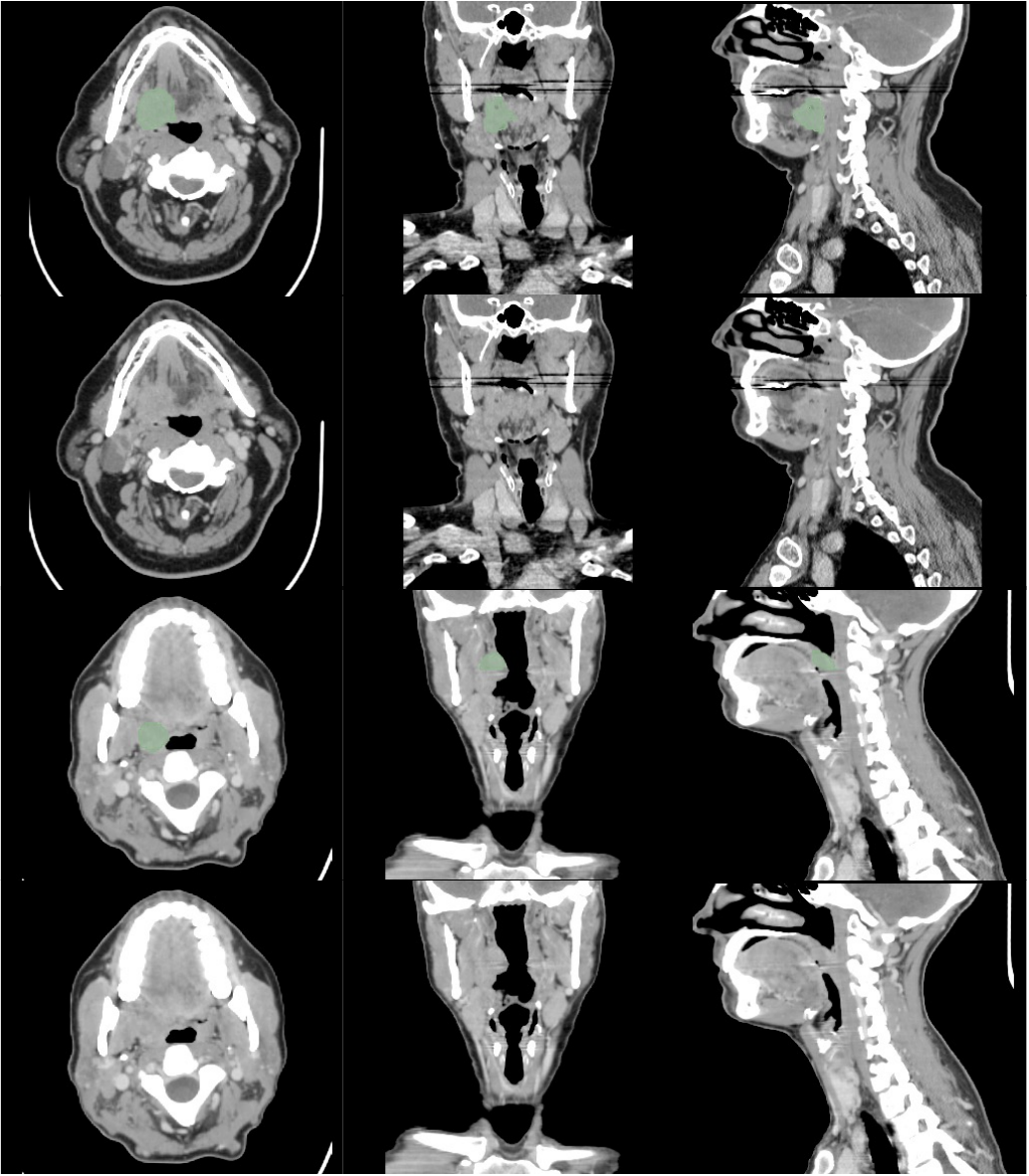}
\caption{Slices in the three axes of the real cases from $CT_{real}^{1258}$ with clipped values for better visualization of the tumor tissue. The tumor is segmented in green.}
\label{fig:app:real_tumour_clip}
\end{figure}

\clearpage
\section{Real MRI scans}
\label{app:Real_MRI_scans}
\begin{figure}[h]
\centering
\includegraphics[width=0.8\linewidth]{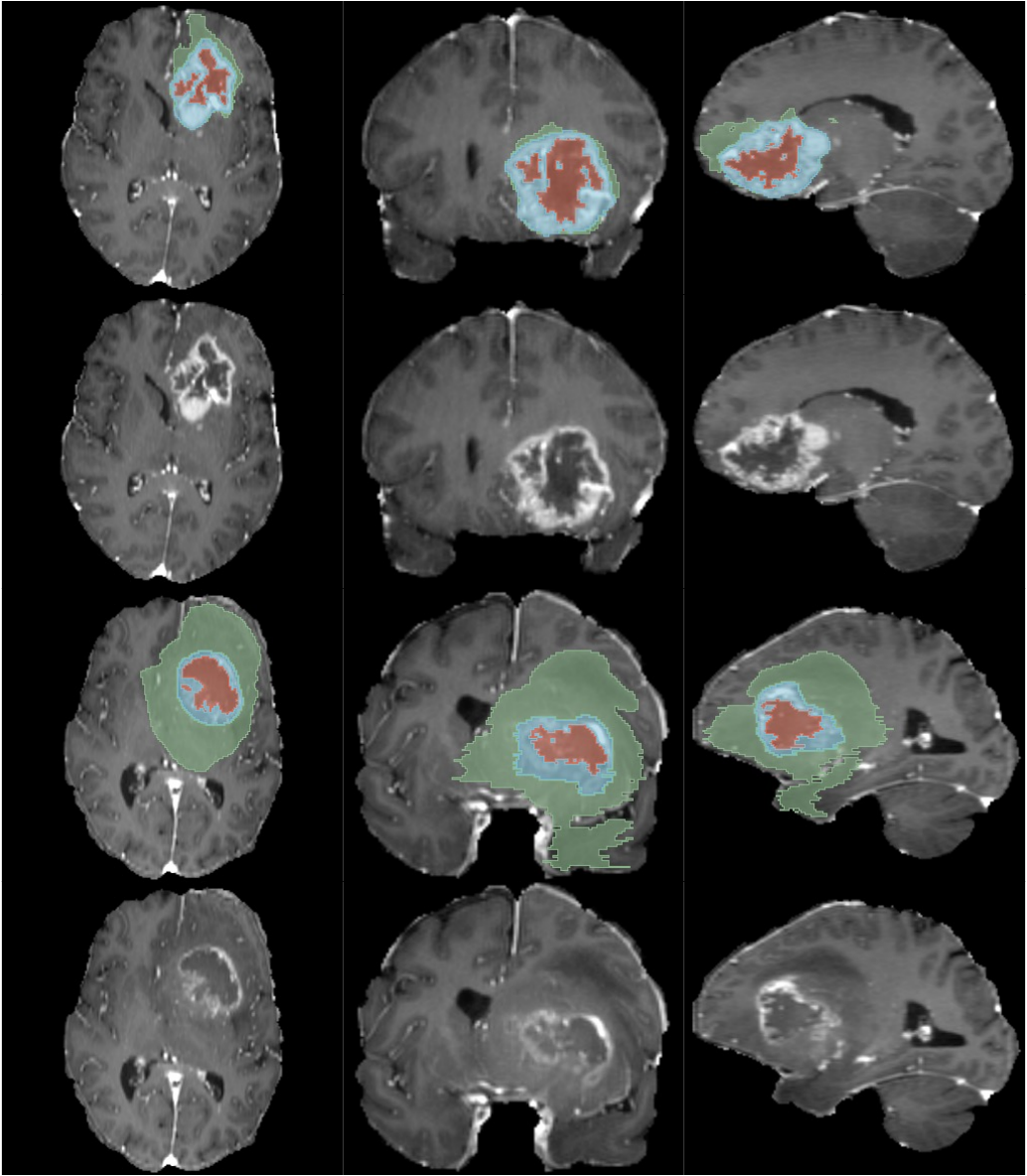}
\caption{Slices in the three axes of the real cases BraTS-GLI-00000-000 (first two rows)  and BraTS-GLI-00002-000 (last two rows). T1c with the necrotic tumor core (red), peritumoral edematous/invaded tissue (green)  and enhancing tumor (blue).}
\label{fig:app:real_MRI}
\end{figure}

\clearpage
\section{cGAN architecture}
\label{app:cGAN_architecture}

\begin{figure}[h]
\centering
\includegraphics[width=1.0\linewidth]{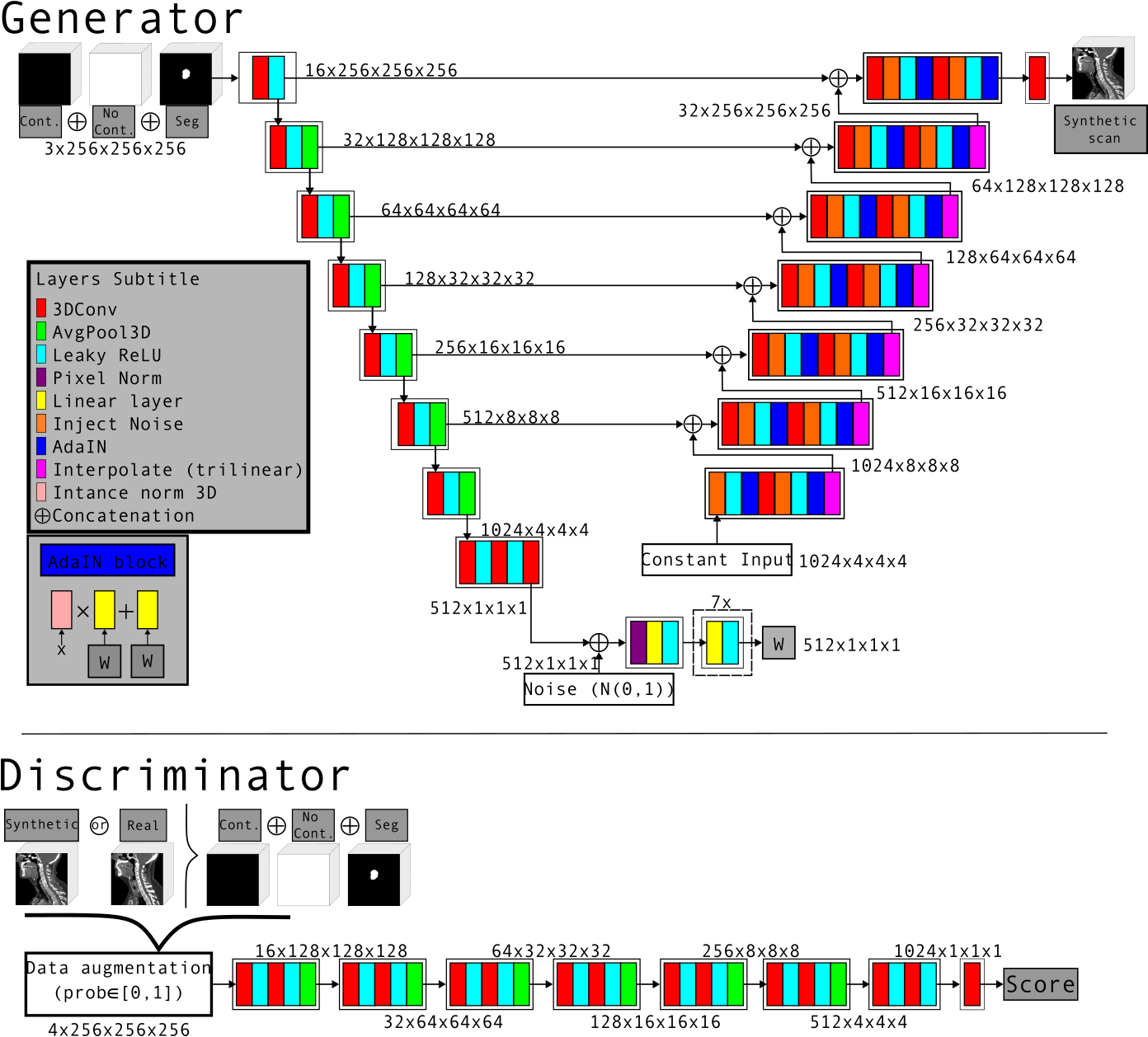}
\caption{Architecture of the cGAN. Generator (top) and Discriminator (bottom).}
\label{fig:cGAN}
\end{figure} 

\clearpage

\clearpage
\section{Inpainting pipeline}
\label{app:inpaint}
\begin{figure}[h]
\centering
\includegraphics[width=0.75\textwidth]{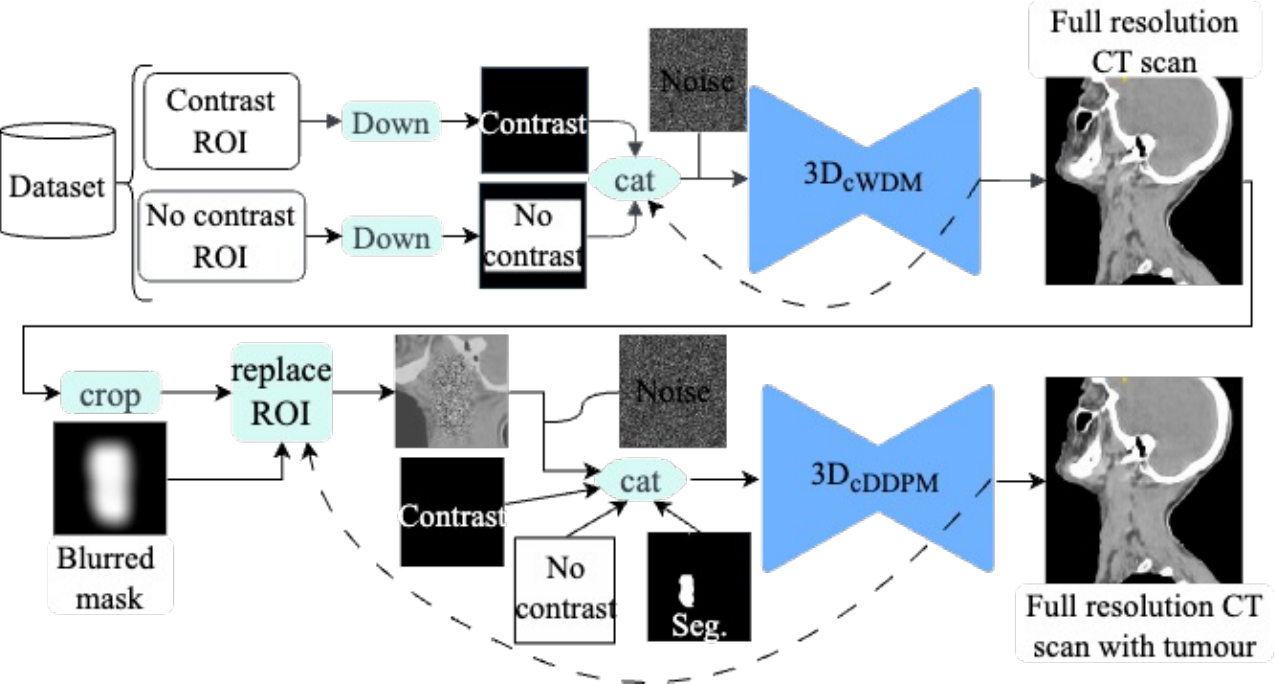}
\caption{ Inpainting pipeline for the generation of a synthetic full resolution CT scan with a tumor. "Down" is the downsample process using nearest neighbor and "cat" the concatenation of the tensors (on the channel dimension).} \label{fig:inpaint}
\end{figure}

\clearpage
\section{Real CT scans with bone segmentation}
\label{app:Real_CT_scans_bone}

\begin{figure}[h]
\centering
\includegraphics[width=0.7\linewidth]{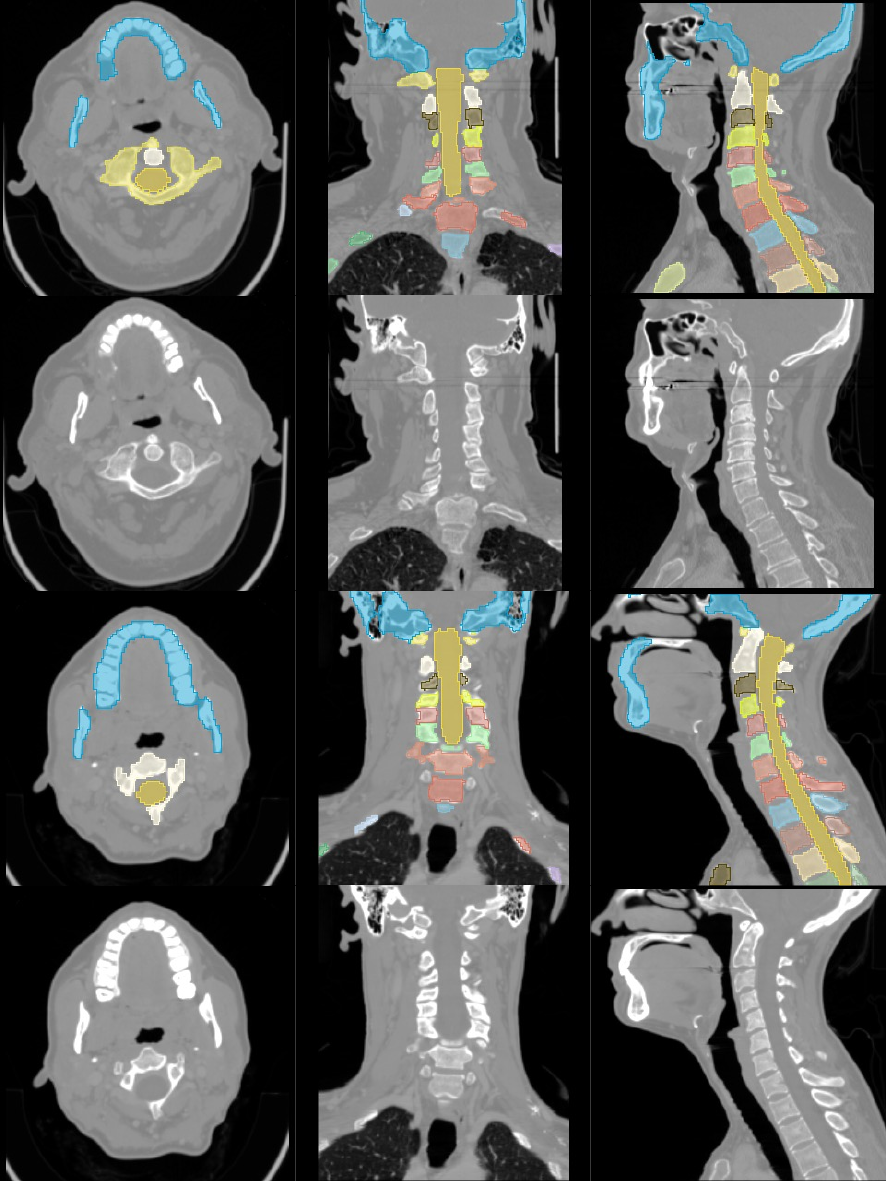}
\caption{Slices in the three axes of real cases from $CT_{real}^{1258}$ and respective predictions of TotalSegmentator.}
\label{fig:app:real_totalseg}
\end{figure}

\begin{figure}[h]
\centering
\includegraphics[width=0.7\linewidth]{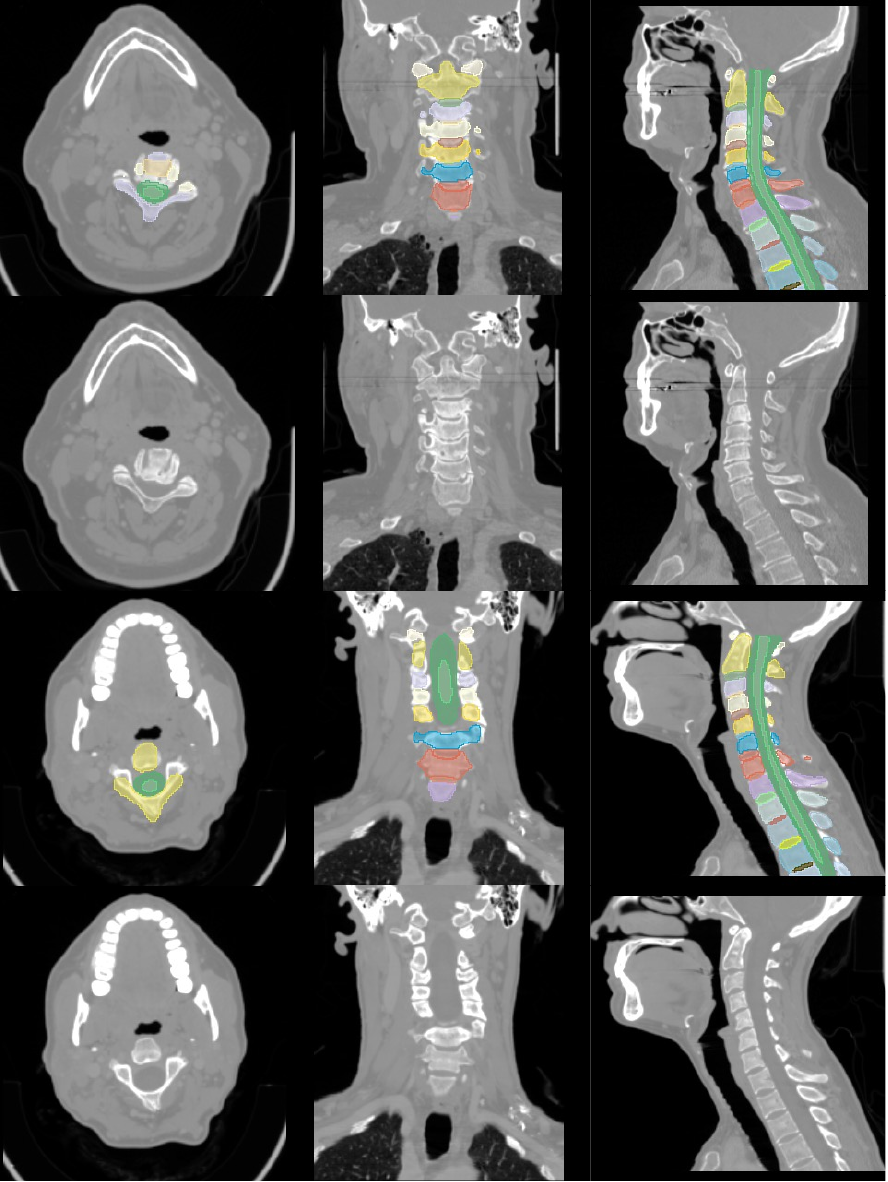}
\caption{Slices in the three axes of real cases from $CT_{real}^{1258}$ and the respective prediction of TotalSpineSeg.}
\label{fig:app:real_spineseg}
\end{figure}

\begin{figure}[h]
\centering
\includegraphics[width=0.7\linewidth]{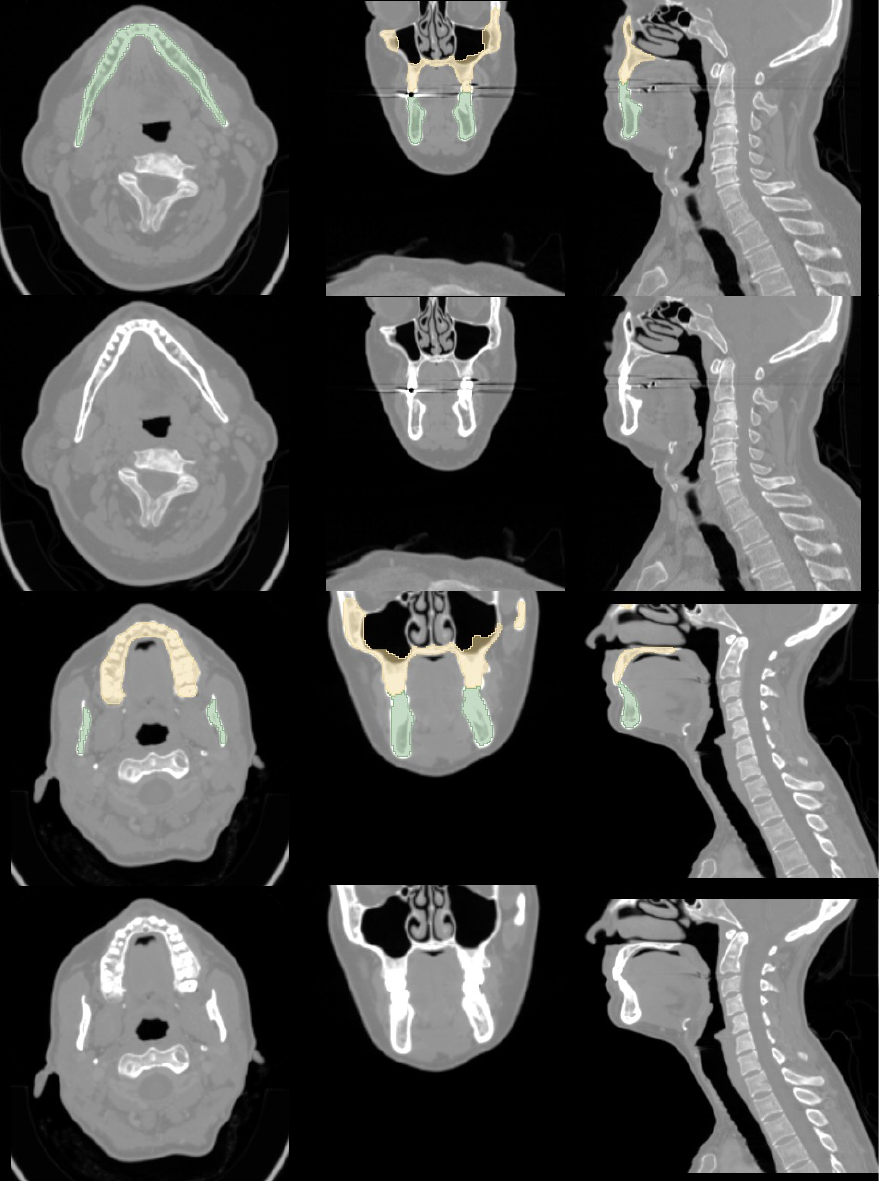}
\caption{Slices in the three axes of real cases from $CT_{real}^{1258}$ and the respective prediction of AMASSS-CBCT.}
\label{fig:app:real_amass}
\end{figure}

\clearpage
\section{Synthetic CT scans with tumor}
\label{app:Synthetic_CT_scans_tumour}

\clearpage
\section{Synthetic MRI scans}
\label{app:Synthetic_MRI_scans}

\clearpage
\section{Synthetic CT scans with bone segmentation}
\label{app:Synthetic_CT_scans_bone}

\clearpage

\end{document}